\documentclass[aps,showpacs,preprintnumbers,amsmath,amssymb]{revtex4}

\usepackage{float}
\usepackage{graphics,epsfig}
\usepackage{graphicx}
\usepackage{dcolumn}
\usepackage{bm}
\usepackage{subeqnarray}
\usepackage{color}
\newcommand{\rmd}{\mathrm{d}}

\begin{document}
\baselineskip=0.8 cm
\title{{\bf Gravitational waves from extreme mass ratio inspirals in Kerr-MOG spacetimes}}

\author{Xiongying Qiao$^{1,2}$, 
        Zhong-Wu Xia$^{1}$, 
        Qiyuan Pan$^{1,3}$\footnote{panqiyuan@hunnu.edu.cn}, 
        Hong Guo$^4$\footnote{hong$\_$guo@usp.br},
        Wei-Liang Qian$^{4,5,3}$\footnote{wlqian@usp.br},
        and Jiliang Jing$^{1,3}$\footnote{jljing@hunnu.edu.cn}}
\affiliation{$^1$Department of Physics, Key Laboratory of Low Dimensional Quantum Structures and Quantum Control of Ministry of Education, Institute of Interdisciplinary Studies, and Synergetic Innovation Center for Quantum Effects and Applications, Hunan Normal University,  Changsha, Hunan 410081, China} 
\affiliation{$^{2}$Department of Physics, Xinzhou Normal University, Xinzhou, Shanxi 034000, China}
\affiliation{$^{3}$Center for Gravitation and Cosmology, College of Physical Science and Technology, Yangzhou University, Yangzhou 225009, China}
\affiliation{$^{4}$Escola de Engenharia de Lorena, Universidade de São Paulo, 12602$-$810, Lorena, SP, Brazil}
\affiliation{$^{5}$Faculdade de Engenharia de Guaratinguet\'a, Universidade Estadual Paulista, 12516-410, Guaratinguet\'a, SP, Brazil}

\vspace*{0.2cm}
\begin{abstract}
\baselineskip=0.6 cm
\begin{center}
{\bf Abstract}
\end{center}

This work elaborates on a detailed analysis of the novel characteristics of gravitational waves (GWs) generated by extreme mass ratio inspirals (EMRIs) within the framework of modified gravity (MOG). 
Our study begins by exploring the geometrical and dynamical properties of the Kerr-MOG spacetime.
We employ the numerical kludge (NK) method for waveform simulations and reveal that the parameter $\alpha$, representing deviations from general relativity (GR), significantly impacts the frequencies of geodesic orbits and, consequently, the EMRI waveforms.
However, the waveform {\it confusion problem} remains mainly unresolved, posing a challenge in distinguishing between the underlying gravitational theories based on the observed EMRI waveforms. Notably, by incorporating the effects of radiation reaction, we observe a substantial reduction in the waveform overlap over time. This reduction could enhance our ability to discern between different waveforms over an extended period.

\end{abstract}
\maketitle
\newpage
\vspace*{0.2cm}

\section{Introduction}

Since the first detection of GWs from a binary black hole merger in 2015~\cite{ligo2016, LIGOScientific:2016vlm}, the GW astrophysics has garnered increasing attention and discussions~\cite{Cai:2017cbj,Chen:2023lsa}.
Among various sources of GW, EMRI systems are of particular interest to future space-based GW detectors~\cite{gair2004, Berry:2019wgg}.
In such an astrophysical system, a small-mass compact object (the secondary with mass $\mu$) orbits and spirals into a central supermassive black hole (the primary with mass $M$), completing hundreds of thousands of orbits before plunging.
During its evolution, which spans tens to hundreds of years, an EMRI system with a mass ratio of $q=\mu/M$ (where $10^{-7}\lesssim q \lesssim 10^{-4}$) emits GWs in the millihertz frequency range~\cite{amaro2017}.
This range is ideally suited for space-based GW detectors such as LISA~\cite{amaro2017, Danzmann:1997hm}, Taiji~\cite{Hu:2017mde, Gong:2021gvw}, and TianQin~\cite{Gong:2021gvw,TianQin:2015yph}.

One of the most significant features of an EMRI system is that the secondary object is gravitationally captured by the central supermassive black hole, whose orbital motion resides in the strong field region of the primary~\cite{Shen:2023pje}. 
Given the secondary object's relatively small mass compared to the primary, it essentially creates a natural spacetime perturbation within such a strong-field region. 
This implies that the GW signals generated by EMRIs carry rich information about near-horizon physics and the strong-field region. Consequently, EMRIs offer a unique method of probing spacetime, allowing for the testing of complex astrophysical environments in galactic nuclei~\cite{Yunes:2011ws, Kocsis:2011dr, Brito:2023pyl, Destounis:2022obl}, the detection of dark matter around supermassive black holes~\cite{Macedo:2013qea, Hannuksela:2019vip, Dai:2023cft}. 
Moreover, EMRIs provide a powerful means of testing GR~\cite{Gair:2012nm,Amaro-Seoane:2019umn,Baibhav:2019rsa}, constraining the nature of Kerr spacetime~\cite{Piovano:2020ooe,Datta:2019euh,Shah:2012gu}. 
The accumulation of a large number of orbital cycles means that even slight deviations from GR can be amplified through long-term GW observations. 
This potential enables the test and restriction of modified gravity theories within the GW spectrum, such as constraints on the Brans-Dicke theory~\cite{Berti:2005qd, Yunes:2011aa}, the detection of dynamical Chern-Simons gravity~\cite{Sopuerta:2009iy, Pani:2011xj, Canizares:2012is}. 
Recently, a series of EMRI studies have focused on detecting scalar charges induced by the dipole radiation~\cite{Maselli:2020zgv, Maselli:2021men}. 
The coupling between scalar fields and spacetime curvature leads to black holes acquiring additional scalar charges, and the research in this area has rapidly expanded to more general scenarios~\cite{Guo:2022euk,Zhang:2021ojz,Barsanti:2022ana, Zhang:2022rfr, Guo:2023mhq, Barsanti:2022vvl,Zi:2022hcc,Liang:2022gdk}.

The present study focuses on the evolution of EMRI within the framework of a central supermassive Kerr-MOG black hole~\cite{moffat2006,Moffat:2014aja}.
The Kerr-MOG black hole model, proposed as an alternative to the Kerr solution in GR, predicts subtle yet profound deviations in the gravitational interaction. 
Notably, extensive research has been conducted on various aspects of Kerr-MOG black holes, such as the observable shadows~\cite{Guo:2018kis,moffat2015,Moffat:2019uxp,Kuang:2022ojj}, superradiance and clouds~\cite{Wondrak:2018fza,Qiao:2020fta}, and merger estimates~\cite{Wei:2018aft}.
Building on this foundation, we extend our analysis to the GW detections from EMRIs of a Kerr-MOG black hole. 
This model provides an essential theoretical framework for testing GR and constraining deviations from Kerr black holes. 

Specifically, this work examines the distinctive EMRI waveforms arising from a Kerr-MOG black hole and explores the implications for GW astrophysics using the NK method~\cite{BabakFGGH}.
Our primary objective is to elucidate the influence of the modified parameter $\alpha$ on the geodesic trajectories and the resulting gravitational waveforms in EMRIs.
Furthermore, we address the {\it confusion problem} that is intrinsic to the waveform analysis.
To be specific, the challenge is posed by the high degree of overlap between Kerr and Kerr-MOG waveforms, which can complicate the discremination between these two spacetime models~\cite{glampedakis2006}.
A pivotal point in our study is the incorporation of radiation reaction effects. 
We demonstrate that the overlap between Kerr and Kerr-MOG waveforms decreases over time when radiation reaction is considered. 
This reduction in the overlap allows for a distinct separation of the waveforms, thereby facilitating the differentiation between these two models of black hole spacetimes.

The structure of the work is organized as follows: 
Section~\ref{chapt:2} delineates the geometric framework of the Kerr-MOG spacetime and the dynamics of geodesic motion, with particular emphasis on the derivation of orbital characteristic frequencies. 
Section~\ref{chapt:3} describes the implementation of the NK method for generating GWs. 
Section~\ref{chapt:4} presents a comprehensive numerical analysis of the trajectory motion and the corresponding EMRI waveforms, addressing the associated confusion problem. 
Finally, Section~\ref{chapt:5} offers conclusions and discussions on the implications of our findings.

\section{Model Construction\label{chapt:2}}

In this section, we elaborate on the EMRI system within the framework of the MOG theory, where the GWs emitted by the binaries will be studied. 
We begin by providing a concise overview of the rotating Kerr-MOG spacetime. Following this, we describe the general orbital trajectories in this context and analytically derive the fundamental frequencies through three integral expressions.

\subsection{Kerr-MOG geometry}

It is well known that the field equations in the MOG theory, also referred to as the scalar-tensor-vector gravity theory~\cite{moffat2006}, are provided by~\cite{moffat2015}:
\begin{align}
    &G_{\mu\nu}=-8\pi G T_{\phi\mu\nu},\label{FieldEquation}\\
    &\nabla_{\nu}B^{\mu\nu} = \frac{1}{\sqrt{-g}}\partial{_\nu}(\sqrt{-g}B^{\mu\nu})=0,\\
    &\nabla_ \sigma B_{\mu \nu}+\nabla_ \mu B_{\nu \sigma}+\nabla_ \nu B_{\sigma \mu}=0,
\end{align}
with the energy-momentum tensor for the vector field $\phi_{\mu}$
\begin{equation}
    T_{\phi\mu\nu}=-\frac{1}{4\pi}\left(B^{\sigma}_{\mu}B_{\nu\sigma}-\frac{1}{4}g_{\mu\nu}B^{\sigma\beta}B_{\sigma\beta}\right),
\end{equation}
where $B_{\mu\nu}=\partial_\mu \phi_\nu-\partial_\nu \phi_\mu$, $G=(1+\alpha)G_N$ with the Newton gravitational constant $G_N$.
The dimensionless deformation parameter $\alpha$ can be used to measure the deviation of MOG from GR.

In Boyer-Lindquist coordinates, the field equation~\eqref{FieldEquation} can be solved to derive a stationary and axisymmetric black hole solution, known as the Kerr-MOG black hole metric~\cite{moffat2015}
\begin{equation}\label{metric}
    ds^2=-\frac{\Delta}{\rho^2 }(cdt-a\sin ^2\theta d\phi )^2+\frac{\rho^2}{\Delta} dr^2+\rho^2d\theta^2+\frac{\sin ^2\theta}{\rho^2}\left [ (r^2+a^2)d\phi-a dt \right ]^2,
\end{equation}
with the definitions
\begin{equation}
    \rho^2\equiv r^2+a^2\cos ^2\theta, ~~~~	\Delta \equiv r^2-2G_NMr+a^2+\frac{\alpha}{(1+\alpha)} G_N^2M^2,
\end{equation}
where $M$ and $a$ are the Arnowitt-Dese-Misner (ADM) mass and spin parameters, respectively. 
Since the charge parameter is proportional to the square root of the modified parameter~\cite{moffat2015}, the physical constraint on the modified parameter satisfies $\alpha\geq0$. 
For simplicity, we adopt $G_N=c=1$ in the remaining paper. 
Similar to the Kerr black hole, the Kerr-MOG metric features two horizons
\begin{equation}
    r_\pm=M\pm\sqrt{\frac{M^2}{1+\alpha}-a^2},
\end{equation}
which revert to the Kerr metric when $\alpha=0$. 
As a result, the ADM mass and spin parameter are constrained by the relation $a/M\leq1/\sqrt{1+\alpha}$.

\subsection{Orbital trajectories}

We begin by examining the geodesic motion of a non-spinning test mass on a bound orbit around a Kerr-MOG black hole. 
Because the metric \eqref{metric} does not explicitly depend on $t$ or $\varphi$, the Kerr-MOG spacetime possesses two Killing vectors: a timelike Killing vector $k^{\mu}=(1,0,0,0)$ and a spacelike killing vector $m^{\mu}=(0,0,0,1)$. 
Consequently, the geodesic motion preserves two conserved quantities
\begin{eqnarray}
    E\equiv-u_t=-g_{tt}u^t-g_{t\phi}u^\phi, ~~~~ L_z\equiv u_\phi=g_{t\phi}u^t+g_{\phi\phi}u^\phi,	
\end{eqnarray}
where $u^{\mu}=\dot{x}^{\mu}=\frac{dx^{\mu}}{d\tau}$ is the four velocity of the secondary with the proper time $\tau$. 
By solving the Hamilton-Jacobi equation, we obtain the equations of geodesic motion
\begin{align}
	&\rho^2 \dot{r}=\pm\sqrt{R},	\label{drdtau}\\
	&\rho^2  \dot{\theta}=\pm\sqrt{\Theta}, \label{dthetadtau}\\
	&\rho^2 \dot{\phi}=\frac{L_z}{\sin^2{\theta}} - aE + \frac{a}{\Delta}\left[  E(r^2+a^2) -aL_z \right], \\
	&\rho^2  \dot{t}=a \left(L_z - aE\sin^2{\theta}\right) + \frac{r^2+a^2}{\Delta} \left[ E(r^2+a^2) - aL_z \right], \label{dtdtau}	
\end{align}
where we have defined
\begin{align}
    &R=\left[ E(r^2 + a^2) - aL_z  \right]^2 - \Delta\left[\mu^2 r^2 + (L_z-aE)^2 + Q\right],\\
    &\Theta=Q -\left[  a^2(\mu^2-E^2) + \frac{L_z^2}{\sin^2{\theta}} \right]\cos^2{\theta},	
\end{align}
with the Carter constant $Q$~\cite{CarterPR1968} and the mass of the secondary~$\mu$.
For the bounded Keplerian geodesics, the orbit can be characterized by three parameters--the eccentricity $e$, the semi latus rectum $p$, and the inclination angle $\iota$, which are defined by
\begin{equation}
    e=\frac{r_a-r_p}{r_a+r_p}, \qquad p=\frac{2r_{a}r_{p}}{r_a+r_p}, \qquad \tan^2\iota=\frac{Q}{L_z^2},
\end{equation}
where $r_a$ and $r_p$ represent the apastron and periastron, respectively.
These are the radial turning points derived from setting the radial velocity derivative $\dot{r}=0$ as delineated in Eq.~\eqref{drdtau}.  
Analogously, for the polar equation of motion~\eqref{dthetadtau}, the polar turning point $\theta_{min}$ can also be obtained by the condition where the time derivative vanishes, i.e., $\dot{\theta}=0$. To circumvent numerical difficulties frequently associated with these turning points, following the method outlined in Ref.~\cite{BabakFGGH}, we transition the integration variables $r$ and $\theta$ to angular coordinates $\psi$ and $\chi$
\begin{equation}
    r=\frac{p}{1+e\cos\psi}, \qquad \cos^{2}\chi=\frac{\cos^{2}\theta}{\cos^{2}\theta_{min}},
\end{equation}
which are more well-behaved for solving the geodesic equations, as the new coordinates $\psi$ and $\chi$ monotonically increase over time, leading to more stable and efficient numerical integration~\cite{BabakFGGH}.

The preceding analysis is confined to the purely geodesic motion, disregarding the effects of radiation reactions. 
However, when incorporating the impact of GW radiation reaction, our findings indicate
\begin{align}\label{inspeqns}
    \frac{d E}{d t} &= f_E (a, M,\alpha, \mu, p, e, \iota), \nonumber \\
    \frac{d L_z}{d t} &= f_L (a, M,\alpha, \mu, p, e, \iota), \nonumber \\
    \frac{d Q}{d t} &= f_Q (a, M,\alpha,  \mu, p, e, \iota).
\end{align}
Considering that, to leading order in $a/M$ and in $M/r$, the Lagrangian for the motion of the secondary in the Kerr-MOG metric aligns with that in the Kerr case~\cite{Ryan9396}, namely,
\begin{eqnarray}
    \mathfrak{L}=\frac{\mu}{2}(\dot{r}^{2}+r^{2}\dot{\theta}^{2}+r^{2}\sin^{2}\theta\dot{\phi}^{2}) +\frac{\mu M}{r}-\frac{2\mu a\sin^{2}\theta}{Mr}\dot{\phi}.
\end{eqnarray}
In the analysis presented, we can disregard the modified parameter  $\alpha$ as shown in Eq.~\eqref{inspeqns}. 
Consequently, we apply the identical radiation-reaction formulas for $f_E$, $f_L$, and $f_Q$ to model the orbital evolution in the Kerr-MOG geometry as we would within the framework of GR, detailed previously in Ref.~\cite{flux}. 
Despite omitting the MOG correction to the radiation reaction at leading order, our results demonstrate that the gravitational waveform generated by a Kerr-MOG black hole exhibits significant deviations from that of a conventional Kerr black hole.

\subsection{Fundamental frequencies}

For an EMRI system, the bounded inclined-eccentric orbit is characterized by three fundamental frequencies $\omega_{r}$, $\omega_{\theta}$, and $\omega_{\phi}$ associated with, respectively, the radial, polar, and azimuthal components of orbital motion. 
To determine these frequencies, it is instructive to rewrite the Hamiltonian $H^{(aa)}$ in terms of the action-angle variables $J_{k}$. 
The latter are particular constants of integration deriving from the Hamilton-Jacobi equation, known as action variables~\cite{GoldsteinPS}, which are determined by performing the cyclic integrals over the spatial conjugate momenta in the Boyer-Lindquist coordinate representation
\begin{align}
    J_{r}&=\frac{1}{2\pi}\oint p_{r}\,\rmd r = \frac{1}{2\pi}\oint\frac{\sqrt{R}}{\Delta}\,\rmd r,	\label{Jr}\\
    J_{\theta}&=\frac{1}{2\pi}\oint  p_{\theta}\,\rmd \theta = \frac{1}{2\pi}\oint\sqrt{\Theta}\,\rmd \theta, \label{JTheta}\\
    J_{\phi}&=\frac{1}{2\pi}\oint p_{\phi}\,\rmd \phi = L_{z} . \label{JPhi}
\end{align}
By definition, these quantities must be functions of the three constants of motion $E$, $L_{z}$, and $Q$, as readily illustrated by the azimuthal action variable~\eqref{JPhi}.
Subsequently, the relevant frequencies can be conveniently computed by taking the partial derivatives of the Hamiltonian with respect to the action-angle variables
\begin{equation}\label{omegak}
    \mu\omega_{k} = \frac{\partial H}{\partial J_{k}}^{(aa)},
\end{equation}
which depend on the mass $\mu$. However, since Eqs.~\eqref{Jr} and \eqref{JTheta} in terms of the non-trivial integrals cannot be solved analytically, the radial action variable $J_{r}$ and polar action variable $J_{\theta}$ do not admit the explicit inversions. 
Fortunately, utilizing a method developed by Schmidt~\cite{Schmidt}, we can calculate the frequencies $\omega_{k}$ in Eq.~\eqref{omegak} even without explicit knowledge of the functional form of the Hamiltonian $H^{(aa)}$ if the theorem on implicit functions is employed.

Now, we come to calculate the fundamental frequencies in detail. Considering the momenta defined as $P_{\beta}^{(aa)}=f_{\beta}^{(aa)}(-\mu^{2}/2,E,L_{z},Q)$, we find $P_{0}^{(aa)}=p_{t}=-E$ and $P_{k}^{(aa)}=J_{k}$. 
Denoting the Jacobian matrix of $f$ by $\mathbf{D}f$ and applying the theorem on implicit functions, we obtain $\mathbf{D}f\cdot \mathbf{D}(f^{-1})=\mathbf{D}f\cdot(\mathbf{D}f)^{-1}=I$, provided that $f$ is non-zero and its Jacobian does not vanish~\cite{Hille}. 
Given that $-\mu^2/2$ represents the invariant value of the Hamiltonian, we have $-\mu^2/2=H^{(aa)}(-E,J_{k})=H(-E,J_{k})$, where we use the symbol $H$ to denote the Hamiltonian $H^{(aa)}$ for convenience. Thus, we arrive at
\begin{equation}
  \left(\begin{array}{llll}
    0 & -1 & 0 & 0 \\ \\
    \frac{\partial J_{r}}{\partial H} &
    \frac{\partial J_{r}}{\partial E} &
    \frac{\partial J_{r}}{\partial L_{z}} &
    \frac{\partial J_{r}}{\partial Q} \\ \\
    \frac{\partial J_{\theta}}{\partial H} &
    \frac{\partial J_{\theta}}{\partial E} &
    \frac{\partial J_{\theta}}{\partial L_{z}} &
    \frac{\partial J_{\theta}}{\partial Q} \\ \\
    0 & 0 & 1 & 0
  \end{array}\right) \cdot
  \left(\begin{array}{llll}
    -\frac{\partial H}{\partial E} &
    \frac{\partial H}{\partial J_{r}} &
    \frac{\partial H}{\partial J_{\theta}} &
    \frac{\partial H}{\partial J_{\phi}} \\ \\
    -1 & 0 & 0 & 0 \\ \\
    0 & 0 & 0 & 1 \\ \\
    -\frac{\partial Q}{\partial E} &
    \frac{\partial Q}{\partial J_{r}} &
    \frac{\partial Q}{\partial J_{\theta}} &
    \frac{\partial Q}{\partial J_{\phi}} \\ \\
   \end{array}\right) = I,
\end{equation}
where two rows of the Jacobian matrix are trivial due to the identities $P_{0}^{(aa)}=-E$ and $J_{\phi}=L_{z}$. 
Moreover, this matrix equation can be split into four distinct non-trivial sets of linear equations
\begin{align}
    -&\mathbf{A}\cdot\frac{\partial}{\partial E}
    \left(\begin{array}{llll}  H \\ Q \end{array}\right) =
    \left(\begin{array}{llll}
      2 \int_{r_{a}}^{r_{2}}\frac{\rmd r}{\sqrt{R}}
      \left\{\frac{(r^{2}+a^{2})[(r^{2}+a^{2})E - a L_{z}]}{\Delta} + a(L_{z}-a E)\right\} \\
      2 a^{2} E\int_{\theta_{min}}^{\pi/2}
      \frac{\cos^{2}\theta\rmd \theta}{\sqrt{\Theta}}
    \end{array}\right), \\
    \label{EqsystActr}
    &\mathbf{A}\cdot\frac{\partial}{\partial J_{r}}
    \left(\begin{array}{llll}  H \\ Q \end{array}\right) =
    \left(\begin{array}{llll} 2\pi \\ 0 \end{array}\right), \\
    &\mathbf{A}\cdot\frac{\partial}{\partial J_{\theta}}
    \left(\begin{array}{llll}  H \\ Q \end{array}\right) =
    \left(\begin{array}{llll} 0 \\ \frac{\pi}{2} \end{array}\right), \\
    &\mathbf{A}\cdot\frac{\partial}{\partial J_{\phi}}
    \left(\begin{array}{llll}  H \\ Q \end{array}\right) =
    \left(\begin{array}{llll}
    2 \int_{r_{a}}^{r_{p}}\frac{\rmd r}{\sqrt{R}}
    \left\{\frac{a[(r^{2}+a^{2})E - a L_{z}]}{\Delta}+(L_{z}-a E)\right\}  \\
    2 L_{z}\int_{\theta_{min}}^{\pi/2}
    \frac{\cot^{2}\theta\rmd \theta}{\sqrt{\Theta}}
    \end{array}\right),
\end{align}
with the coefficient matrix
\begin{align}
    \mathbf{A} =
    \left(\begin{array}{llll}
      2\int_{r_{a}}^{r_{p}}\frac{r^{2}\rmd r}{\sqrt{R}} &
      -\int_{r_{a}}^{r_{p}}\frac{\rmd r}{\sqrt{R}} \\ \\
      2 a^{2}\int_{\theta_{min}}^{\pi/2}
      \frac{\cos^{2}\theta\,\rmd \theta}{\sqrt{\Theta}} &
      \int_{\theta_{min}}^{\pi/2}\frac{\rmd \theta}{\sqrt{\Theta}}
    \end{array}\right).
\end{align}
Thus, with the definitions of the radial integrals
\begin{align}
    \label{EqRadIntegrx}
    X(r_{a},r_{p}) &= \int_{r_{a}}^{r_{p}}\frac{\rmd r}{\sqrt{R}}, \\
    \label{EqRadIntegry}
    Y(r_{a},r_{p}) &= \int_{r_{a}}^{r_{p}}\frac{r^{2}\,\rmd r}{\sqrt{R}}, \\
    \label{EqRadIntegrz}
    Z(r_{a},r_{p}) &= \int_{r_{a}}^{r_{p}}
    \frac{dr}{\Delta\sqrt{R} \left(1+\alpha \right)} \big \{ a M E \left [2 (\alpha +1) r-\alpha  M\right ]+\alpha  L_{z} (M-r)^2+L_{z} r (r-2 M)\big \},\\
    \label{EqRadIntegrw}
    W(r_{a},r_{p}) &=
    \int_{r_{a}}^{r_{p}}\frac{dr}{\Delta\sqrt{R}}\left[a^2 E \left(r^2+2 M r-\frac{\alpha  M^2}{\alpha +1}\right)+a L_z M \left(\frac{\alpha  M}{\alpha +1}-2 r\right)+r^4 E \right],
\end{align}
we obtain the solutions of the above systems of equations
\begin{align}
    \label{EqhDervp0}
    -\frac{\partial H}{\partial E} &=
    \frac{K(k)W(r_{a},r_{p}) +
        a^{2}z_{+}^{2}E\left[K(k)-E(k)\right]X(r_{a},r_{p})
        }{K(k)Y(r_{a},r_{p}) +
          a^{2}z_{+}^{2}\left[K(k)-E(k)\right]X(r_{a},r_{p})}, \\
    \label{EqhDervjr}
    \frac{\partial H}{\partial J_{r}}  &=
    \frac{\pi K(k)}{K(k)Y(r_{a},r_{p}) + a^{2}z_{+}^{2}\left[K(k)-E(k)\right] X(r_{a},r_{p})}, \\
    \label{EqhDervjtheta}
    \frac{\partial H}{\partial J_{\theta}}  &=
    \frac{\pi\beta z_{+}X(r_{a},r_{p})}{2\{K(k)Y(r_{a},r_{p}) +
        a^{2}z_{+}^{2}\left[K(k)-E(k)\right]X(r_{a},r_{p})\}}, \\
    \label{EqhDervjphi}
    \frac{\partial H}{\partial J_{\phi}}  &=
    \frac{K(k)Z(r_{a},r_{p}) + L_{z}[\Pi(z_{-}^{2},k)-K(k)]X(r_{a},r_{p})
        }{K(k)Y(r_{a},r_{p}) + a^{2}z_{+}^{2}[K(k)-E(k)]X(r_{a},r_{p})},
\end{align}
with
\begin{eqnarray}\label{kbeta}
    k=z_{-}^{2}/z_{+}^{2}, \qquad \beta^{2}=a^{2}(\mu^{2}-E^{2}),
\end{eqnarray}
where $z_{\pm}^{2}$ represent the two roots of the equation $\Theta(z)=0$, derived by substituting $z=\cos\theta$ into $\Theta$. 
Furthermore, $K(k)$, $E(k)$ and $\Pi(z_{-}^{2},k)$ are, respectively, the complete elliptical integrals of the first, second, and third kind
\begin{align}
    \label{eq:ell_integr_k}
    K(k) &=\int_{0}^{\pi/2}\frac{\rmd \psi}{\sqrt{1-k\sin^{2}\psi}}, \\
    \label{eq:ell_integr_e}
    E(k) &=\int_{0}^{\pi/2}\sqrt{1-k\sin^{2}\psi}\,\rmd \psi, \\
    \label{eq:ell_integr_pi}
    \Pi({z_{-}^{2},k}) &=
    \int_{0}^{\pi/2}\frac{\rmd\psi}{\left(1-z_{-}^{2}\sin^{2}\psi\right)
    \sqrt{1-k\sin^{2}\psi}}.
\end{align}
Utilizing Eq.~\eqref{omegak}, we finally have
\begin{align}
    \label{EqFreqr}
    \omega_{r} &= \frac{\pi K(k)}{(Y+a^{2}z_{+}^{2}X)K(k)-a^{2}z_{+}^{2}X E(k)}, \\
    \label{EqFreqtheta}
    \omega_{\theta} &= \frac{\pi\beta z_{+}X}{2[(Y+a^{2}z_{+}^{2}X)K(k)-a^{2}z_{+}^{2}X E(k)]}, \\
    \label{EqFreqphi}
    \omega_{\phi} &= \frac{(Z-L_{z}X)K(k) + L_{z}X\Pi(z_{-}^{2},k)}{(Y+a^{2}z_{+}^{2}X)K(k)-a^{2}z_{+}^{2}X E(k)},
\end{align}
which depend on the background parameters $(M, a, \alpha)$, orbital parameters $(e, p, \iota)$ and secondary parameter $\mu$ simultaneously.

\begin{figure}[ht]
\includegraphics[scale=0.32]{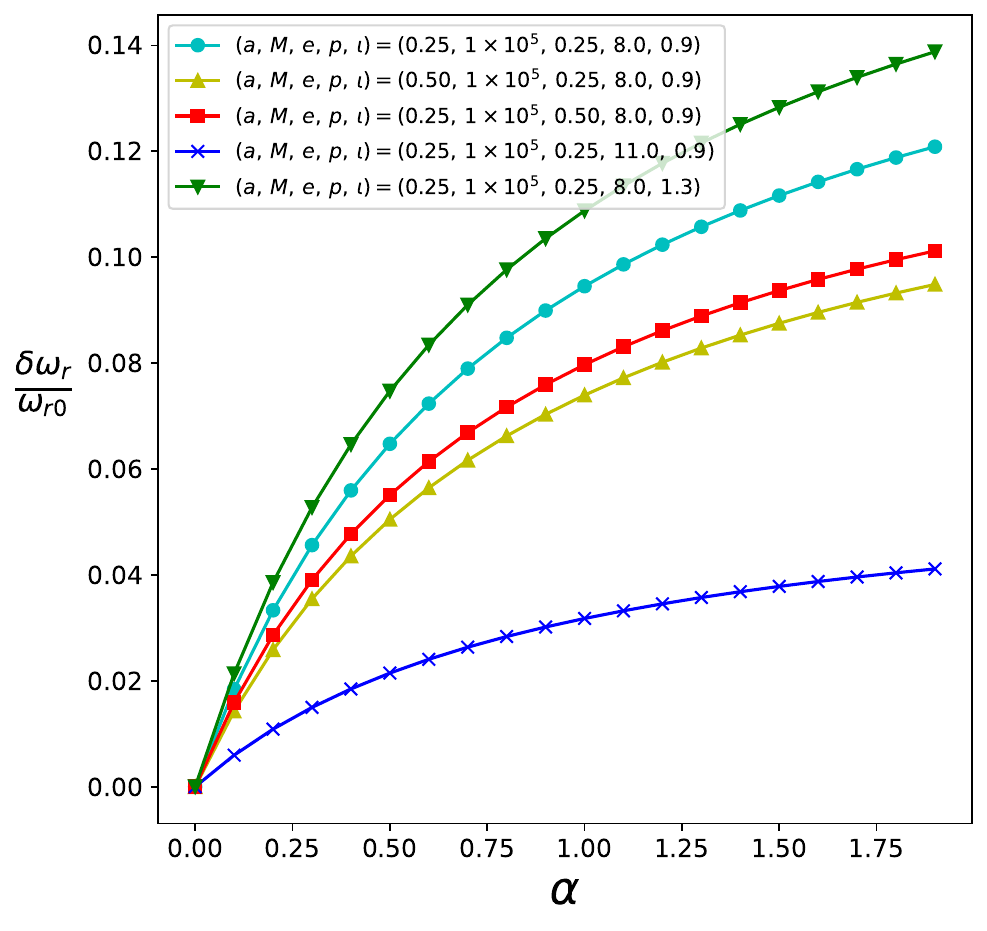}\hspace{0.2cm}%
\includegraphics[scale=0.32]{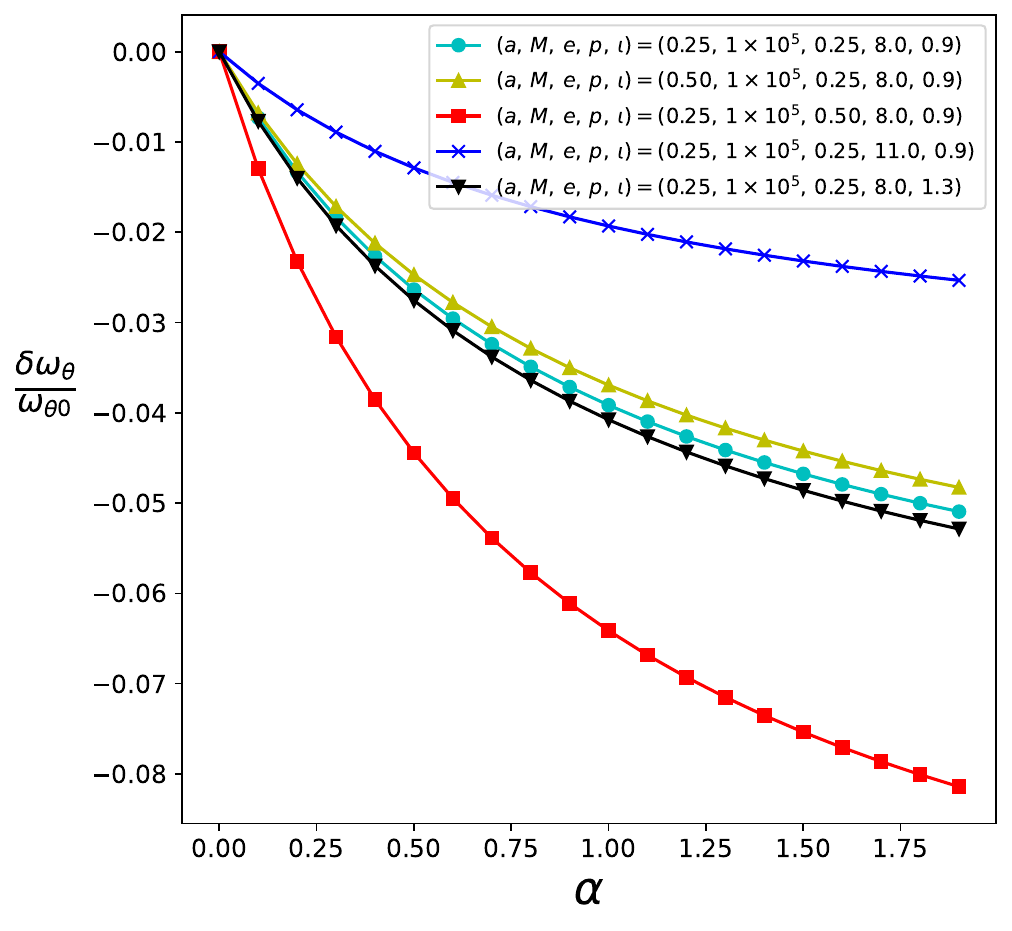}\hspace{0.2cm}%
\includegraphics[scale=0.32]{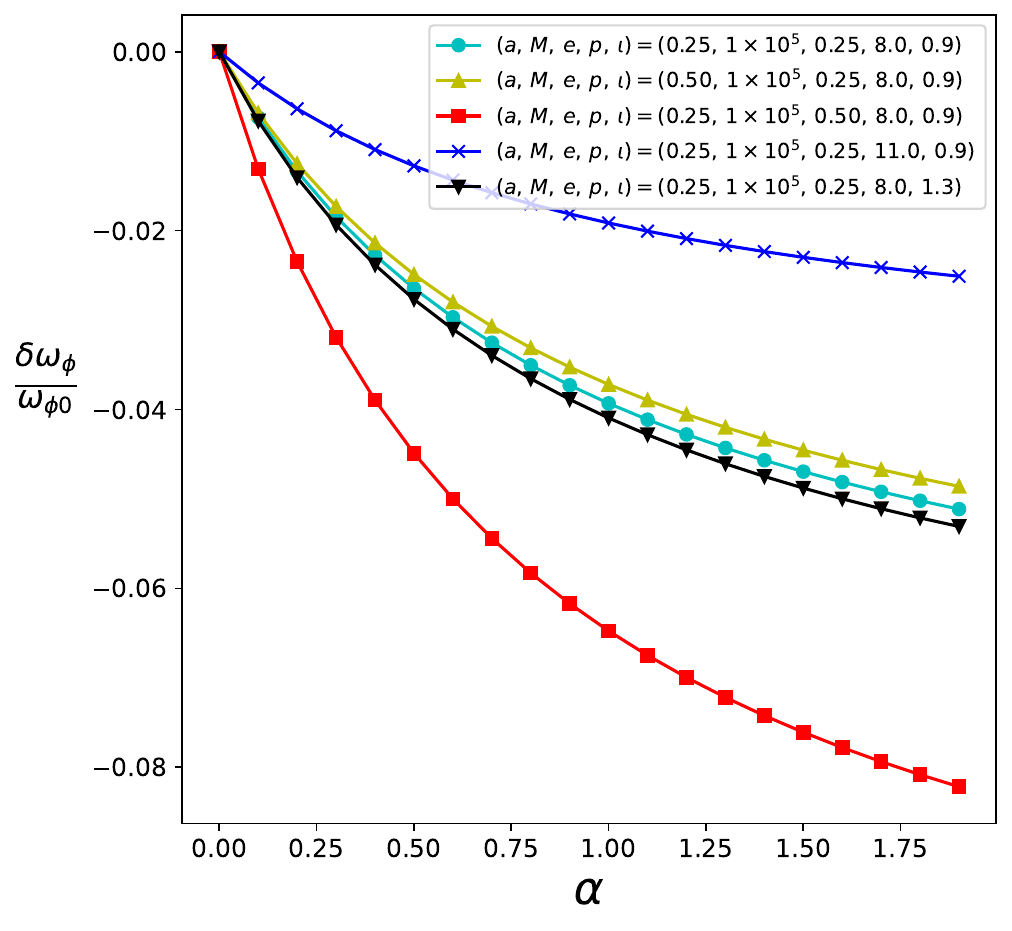}\\ \vspace{0.0cm}
\caption{\label{fig:alpha_3D_freq} (Color online) Frequency shifts as functions of the parameter $\alpha$ in case of various orbits/black hole parameters $(a, M, e, p, \iota)$. We use $\omega_0$ to denote the frequencies evaluated at $\alpha=0$.}
\end{figure}

To directly observe the influence of the modified parameter $\alpha$ on the fundamental frequencies, we present in Fig.~\ref{fig:alpha_3D_freq} the frequency shifts $\Delta\omega/\omega_0 = (\omega - \omega_0)/\omega_0$ as a function of the modified parameter $\alpha$ while holding the parameters $(a, M, e, p, \iota)$ constant, where $\omega_0$ denotes the frequency of the Kerr geometry and we set the secondary mass $\mu=1M_\odot$. Notably, with the increase of the modified parameter $\alpha$, there is a discernible increase in the radial frequency shift, whereas the azimuthal and polar frequency shifts exhibit a decrease. 
This pattern indicates that the augmentation of $\alpha$ primarily elevates the radial frequency $\omega_r$ but reduces the polar frequency $\omega_{\theta}$ and azimuthal frequency $\omega_{\phi}$.

\section{Waveform modeling and signal analysis\label{chapt:3}}

Next, we compute the gravitational waveforms by employing the particle's trajectory in the Kerr-MOG geometry and the kludge waveform generation method. 
Following the approach outline in Ref.~\cite{BabakFGGH}, we introduce the flat-space Cartesian coordinates
\begin{align}
    x& = r\sin\theta\cos\phi, \nonumber \\
    y& = r\sin\theta\sin\phi, \nonumber \\
    z& = r\cos\theta,
\end{align}
and employ the multipolar expansion of metric perturbations that characterize the GWs emitted by an isolated system~\cite{ThorneRMP}. 
To circumvent the intricacies of extensive computations while capturing the essential physics, we focus on the lowest-order term and employ the quadrupole formula
\begin{eqnarray}
    \bar{h}^{ij}(t,\mathbf{x})&=&\frac{2}{R}\left[\ddot{I}^{ij}(t^{\prime})\right]_{t^{\prime}=t-r},
\end{eqnarray}
with the source's mass quadrupole moment
\begin{eqnarray}
    I^{ij}(t^{\prime}) &=& \mu x_{p}^{\prime i}x_{p}^{\prime j},
\end{eqnarray}
where $R$ denotes the luminosity distance from the source to the observer, and the dot denotes the derivative with respect to the coordinate time $t$. 
Following the procedure proposed in Ref.~\cite{BabakFGGH}, we apply the transverse-traceless (TT) projection to the previously mentioned expressions, thereby deriving the waveform in the standard TT gauge, i.e., the plus and cross components of the waveform at the observation point's latitudinal angel $\Theta$ and azimuthal angel $\Phi$~\cite{XinHanYangPRD2019}
\begin{align}
    h_+ =& \left\{ \cos^2 \Theta \left[h^{xx} \cos^2 \Phi + h^{xy} \sin(2\Phi)+h^{yy} \sin^2 \Phi\right] + h^{zz} \sin^2\Theta - \sin(2\Theta)(h^{xz} \cos \Phi + h^{yz} \sin \Phi) \right\}\nonumber \\
        &-\left[h^{xx} \sin^2\Phi - h^{xy} \sin (2\Phi) + h^{yy} \cos^2 \Phi\right],\nonumber \\
    h_\times =&  2 \left\{ \cos\Theta\left[-\frac 12 h^{xx} \sin (2\Phi) + h^{xy} \cos (2\Phi) + \frac 12 h^{yy} \sin (2\Phi) \right] + \sin \Theta\left(h^{xz} \sin \Phi - h^{yz} \cos \Phi\right)\right\},
\end{align}
which can be used to construct the response of the LISA detectors.

In practical EMRI data analysis, the detector's response to an incoming GW can be represented as~\cite{Canizares:2012is}
\begin{eqnarray}
    h_{\beta}(t)=\frac{\sqrt{3}}{2}\left[F^{+}_{\beta}(t)h^{}_{+}(t)+F^{\times}_{\beta}(t)h^{}_{\times}(t)\right]\,,
\end{eqnarray}
with an index $\beta$ denoting the distinct, independent channels of the detector. 
For the LISA detector, two independent Michelson-like interferometer channels are constructed from the data stream. 
This dual-channel approach enables us to capture two distinct waveforms of the GWs, thereby enhancing the validity of our conclusions.
Following the approach detailed in~\cite{BarackCutler}, the antenna pattern functions, which describe the detector's responses, are 
\begin{align}
    F^{+}_I &= \frac{1}{2}(1+\cos^2\Theta)\cos(2\Phi)\cos(2\psi) -\cos\Theta\sin(2\Phi)\sin(2\psi)\,,  \nonumber \\
    F^{\times}_I &= \frac{1}{2}(1+\cos^2\Theta)\cos(2\Phi)\sin(2\psi)+\cos\Theta\sin(2\Phi)\cos(2\psi)\,, \nonumber \\
    F^{+}_{II} &= \frac{1}{2}(1+\cos^2\Theta)\sin(2\Phi)\cos(2\psi) +\cos\Theta\cos(2\Phi)\sin(2\psi)\,, \nonumber \\
    F^{\times}_{II} &= \frac{1}{2}(1+\cos^2\Theta)\sin(2\Phi)\sin(2\psi) -\cos\Theta\cos(2\Phi)\cos(2\psi)\,,
\end{align}
with the angles $(\Theta(t),\Phi(t))$~\cite{Cutler1998}
\begin{align}
    \cos\Theta^{}(t) &= \frac{1}{2}\cos\theta^{}_{\rm S}-\frac{\sqrt{3}}{2}\sin\theta^{}_{\rm S}\cos\left(\frac{2\pi t}{T}-\phi^{}_{\rm S}\right) \,,\nonumber \\
    \Phi^{}(t)& = \frac{2\pi t}{T}+\tan^{-1}\left[\frac{\sqrt{3}\cos\theta^{}_{\rm S}+\sin\theta^{}_{\rm S}\cos\left(2\pi t/T -\phi^{}_{\rm S}\right)}{2\sin\theta^{}_{\rm S}\sin\left(2\pi t/T -\phi^{}_{\rm S}\right)}\right]\,,
\end{align}
and the polarization angle $\psi$
\begin{align}
    \tan \psi =& \left[ \left\{ \cos\theta^{}_{\rm K} -
        \sqrt{3}\sin\theta^{}_{\rm K}\cos\left(2\pi t/T -\phi^{}_{\rm K}\right)\right\}-2\cos\Theta(t)\left\{\cos\theta^{}_{\rm K}\cos\theta^{}_{\rm S} \right.\right. \nonumber\\ 
        &\left.\left.+\sin\theta^{}_{\rm K}\sin\theta^{}_{\rm S}\cos(\phi^{}_{\rm K}-\phi^{}_{\rm S}) \right\} \frac{}{}\right]/\left[\frac{}{} \sin\theta^{}_{\rm K}\sin\theta^{}_{\rm S}\sin(\phi^{}_{\rm K}-\phi^{}_{\rm S})\right.\nonumber\\ 
        &\left. -\sqrt{3}\cos\left(2\pi t/T \right)\left\{\cos\theta^{}_{\rm K}\sin\theta^{}_{\rm S}\sin\phi^{}_{\rm S}-\cos\theta^{}_{\rm S}\sin\theta^{}_{\rm K}\sin\phi^{}_{\rm K} \right\} \right.\nonumber\\
        &\left. -\sqrt{3}\sin\left(2\pi t/T\right) \left\{\cos\theta^{}_{\rm S}\sin\theta^{}_{\rm K}\cos\phi^{}_{\rm K}-\cos\theta^{}_{\rm K}\sin\theta^{}_{\rm S}\cos\phi^{}_{\rm S} \right\} \frac{}{}\right],\label{LISApsi}
\end{align}
where the spherical polar angles $(\theta^{}_{\rm S},\phi^{}_{\rm S})$ describe the sky location of the EMRI concerning the Solar System barycenter (SSB) frame, and the angles $(\theta^{}_{\rm K},\phi^{}_{\rm K})$ determine the direction of the massive black hole spin with respect to the SSB frame, respectively. 
It is pertinent to mention that $T=1$ yr represents the period of the Earth's orbit around the Sun.

On the other hand, to analyze the distinctions between waveforms in Kerr-MOG spacetimes and those predicted from GR, we employ the NK method and utilize the overlap function to assess differences as proposed by Glampedakis \emph{et al.}~\cite{glampedakis2006}. 
For the two time-series signals, namely a signal $h_1(t)$ and a template $h_2(t)$, we can define the internal product in terms of the Fourier transforms indicated by a ``tilde''~\cite{cutler, BalasubramanianSD}
\begin{equation}
    \left( h_1 , h_2 \right) \equiv 2 \int_0^{\infty} \frac{{\tilde h}_1^*(f) {\tilde h}_2(f) + {\tilde h}_1(f) {\tilde h}_2^*(f)}{S_n(f)}df\;,
\end{equation}
where $S_{n}(f)$ is the spectral sensitivity for the LISA detector~\cite{LISAPSD}. 
Thus, the overlap function $\cal O$ is defined as
\begin{equation}
	{\cal O}(h_1,h_2)\equiv  \frac{(h_1,h_2)}{(h_1,h_1)^{1/2}(h_2, h_2)^{1/2}}\;,
\end{equation}
where ${\cal O}(h_1,h_2)=1$ if the two waveforms are identical, ${\cal O}(h_1,h_2)=0$ if they are
totally uncorrelated, and ${\cal O}(h_1,h_2)=-1$ if they are perfectly anticorrelated~\cite{BarausseRPA}.


\section{Numerical Results}\label{chapt:4}

Now, we are equipped to explore the EMRIs in Kerr-MOG spacetimes using the NK waveforms~\cite{BabakFGGH}. In the following numerical calculations, we set the primary mass as $M=1\times 10^6M_\odot$ and the secondary mass as $\mu=10M_\odot$, resulting in a mass ratio of $q=\mu/M=10^{-5}$. The remaining parameters are the same as those given in~\cite{Canizares:2012is}, namely, $\theta_S=\phi_S=1.57$, $\theta_K=0.329$, $\phi_K=0.78$, and the luminosity distance is $D_L=1\mathrm{Gpc}$. To assess the influence of the modified parameter $\alpha$ on EMRIs, we first focus on the equatorial geodesics while considering only the orbital evolutions dictated by the conservative dynamics without explicitly taking into account the radiation reaction. We address the issue of waveform confusion~\cite{glampedakis2006}, which arises when attempting to differentiate between Kerr-MOG and Kerr signals.
In this regard, the radiation reaction is incorporated into the formulism partly to resolve this difficulty.
Subsequently, we extend our study to examine the impact of $\alpha$ on the EMRI waveforms for inclined orbits.

\subsection{Orbital evolution and EMRI waveforms}

\begin{figure}[ht]
\includegraphics[scale=0.35]{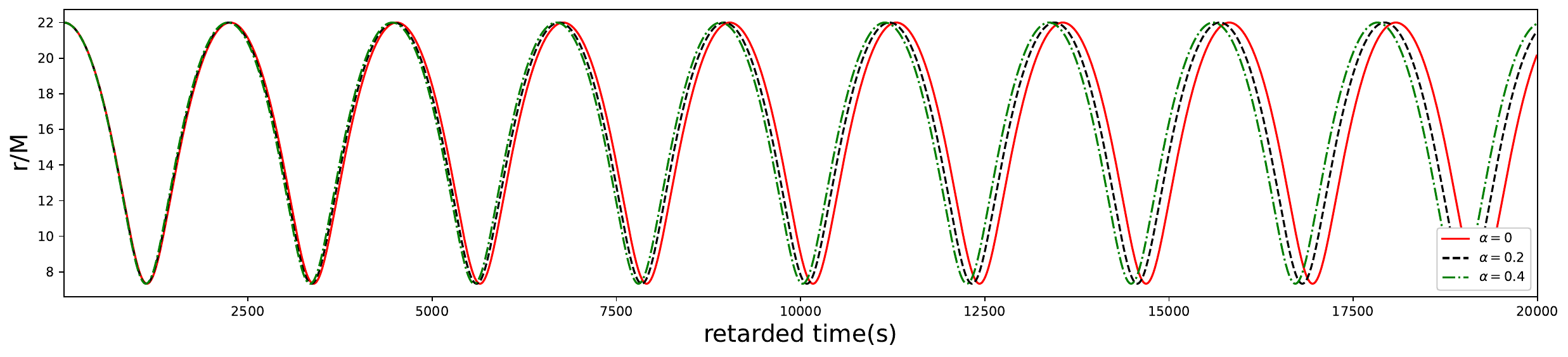}\\ \vspace{0.0cm}
\includegraphics[scale=0.35]{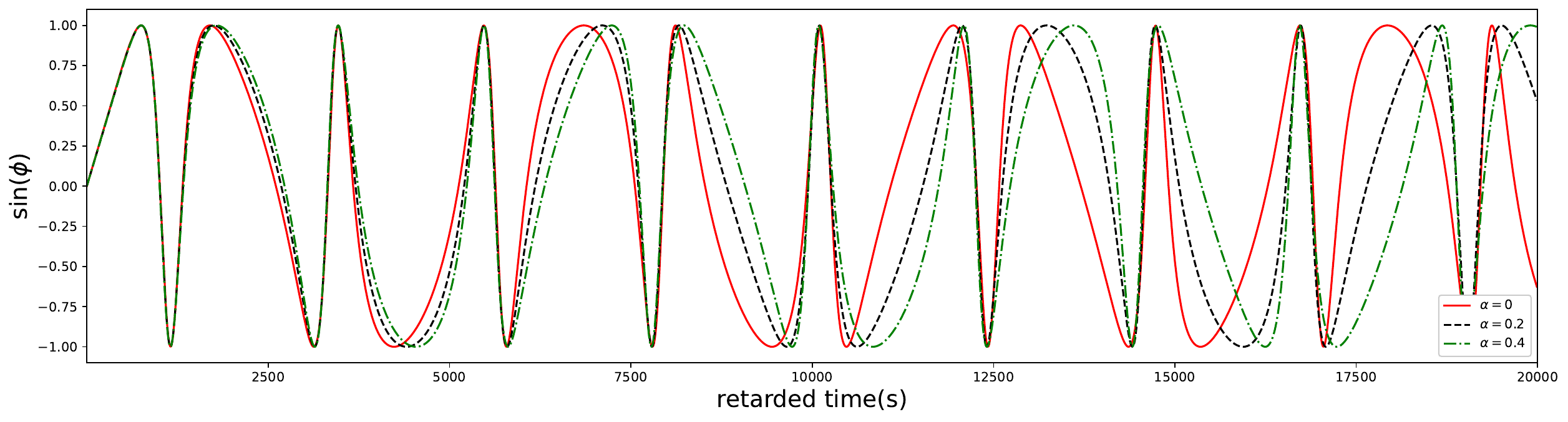}\\ \vspace{0.0cm}
\caption{\label{EquatorialOrbitRPhi} (Color online) Time series of motion in the $r$ and $\phi$ directions of the Kerr-MOG orbit with various parameters $\alpha$ for the first 20000 s in the case of $(a,M,e,p)=(0.25,1\times 10^{6} M_\odot,0.5,11.0)$. 
The red, black, and green lines in each panel correspond to increasing modified parameters, i.e., $\alpha=0$, $0.2$, and $0.4$, respectively.}
\end{figure}

\begin{figure}[ht]
\includegraphics[scale=0.35]{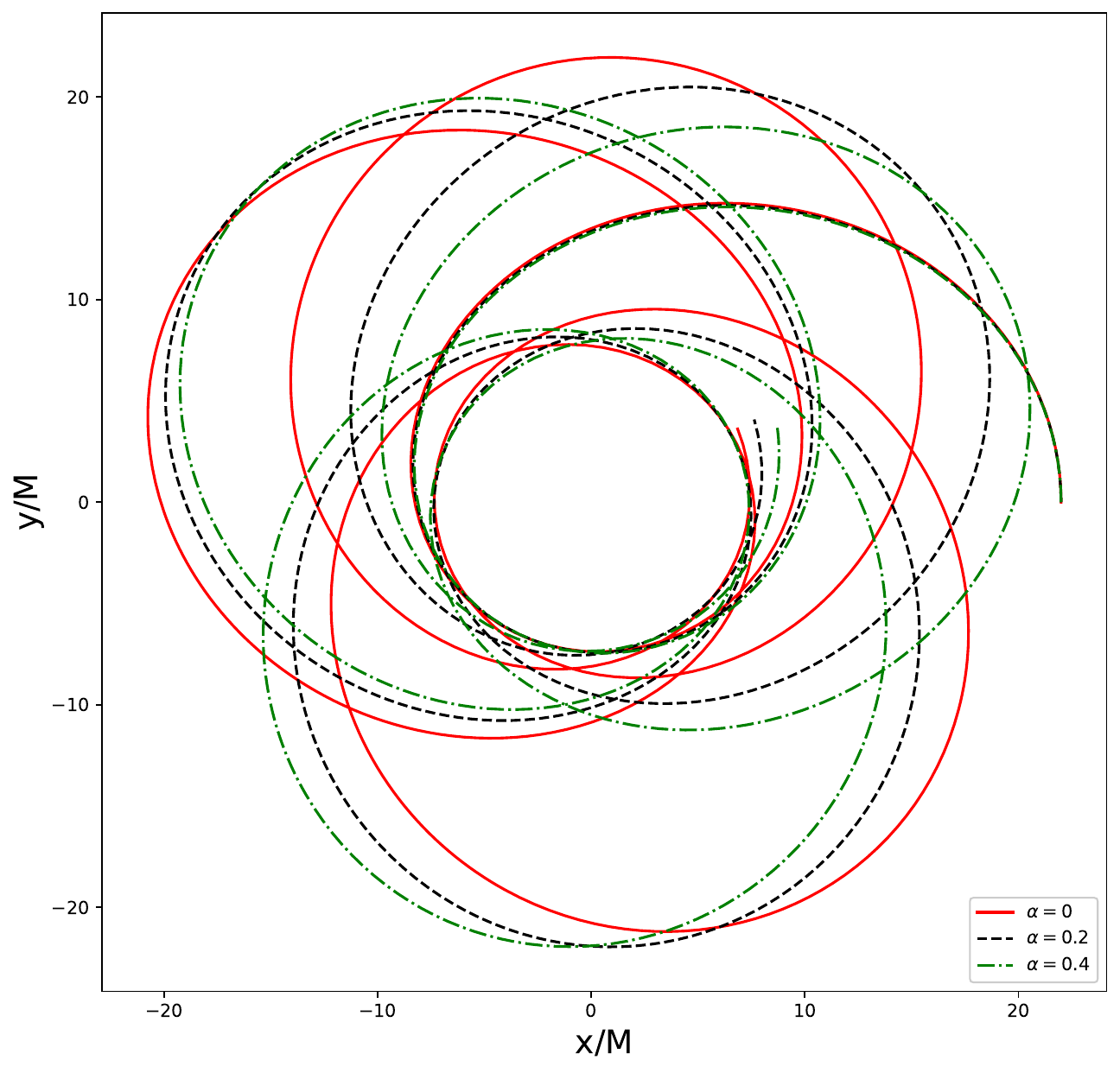}\\ \vspace{0.0cm}
\includegraphics[scale=0.35]{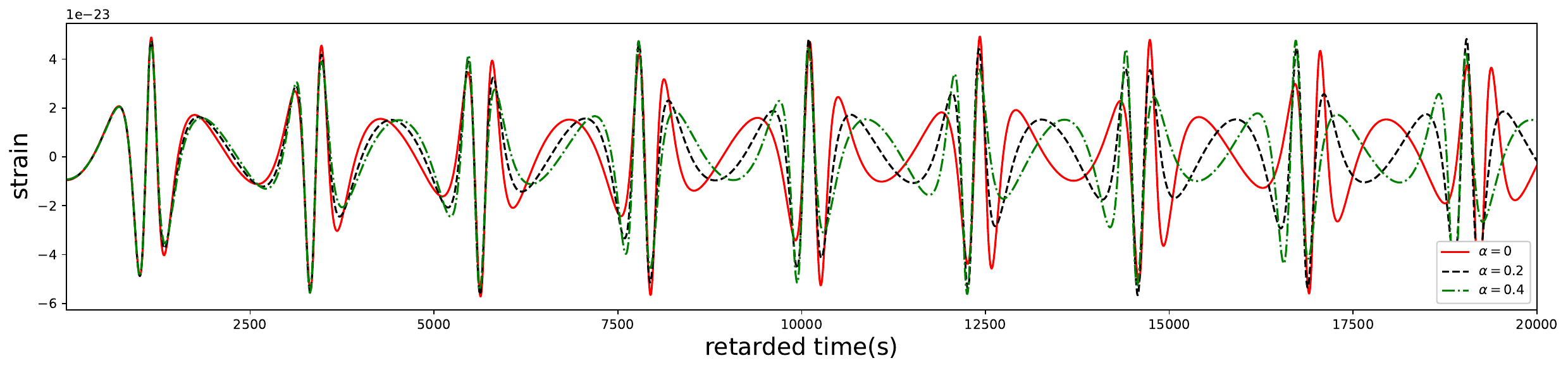}\\ \vspace{0.0cm}
\caption{ \label{EquatorialOrbitGW} (Color online) Orbits and the component $h_I$ of the EMRI waveforms of the Kerr-MOG orbit with various parameters $\alpha$ in the case of $(a, M,e,p)=(0.25,1\times 10^{6} M_\odot,0.5,11.0)$. 
The red, black, and green lines in each panel correspond to increasing modified parameters, i.e., $\alpha=0$, $0.2$, and $0.4$, respectively.}
\end{figure}

For the equatorial geodesic orbits of the conservative dynamics, the initial values of $(E, L_z, Q)$ are given by the following equations~\cite{AK}
\begin{equation}
    R(E,L_z,Q=0,r=r_a)=0, ~~~~R(E,L_z,Q=0,r=r_p)=0, ~~~~Q=0.
\end{equation}
For the same initial positions and parameters $e$ and $p$, in Fig.~\ref{EquatorialOrbitRPhi}, we present the time series of motion in the $r$ and $\phi$ directions for a Kerr-MOG orbit characterized by various values of the modified parameter $\alpha$ over the first 20000 seconds.
We observe that  $\alpha$ significantly influences the orbit phase; meanwhile, its impact on the amplitude is minimal.
As $\alpha$ increases, the period in the $r$ direction decreases, whereas the period in the $\phi$ direction increases. 
This observation aligns with the results shown in Fig.~\ref{fig:alpha_3D_freq}, which demonstrates that an increase in $\alpha$ elevates the radial frequency $\omega_r$ and reduces the polar frequency $\omega_\phi$.
The effects of $\alpha$ on the trajectories become particularly apparent after several orbital cycles. 
With the geodesic orbits, we can generate GWs using the kludge method.
As illustrated in Fig.~\ref{EquatorialOrbitGW}, the equatorial trajectory over the initial 8000 seconds and corresponding component $h_I$ of the EMRI waveform over the first 20000 seconds are markedly affected by  $\alpha$.
These findings enable us to distinguish the GW signals of Kerr-MOG black holes from those of  Kerr black holes.
It is important to note that although we have only presented results for a specific set of $(a, M,e,p)$, our conclusions are valid for all other $(a, M,e,p)$.

\begin{figure}[ht]
\includegraphics[scale=0.35]{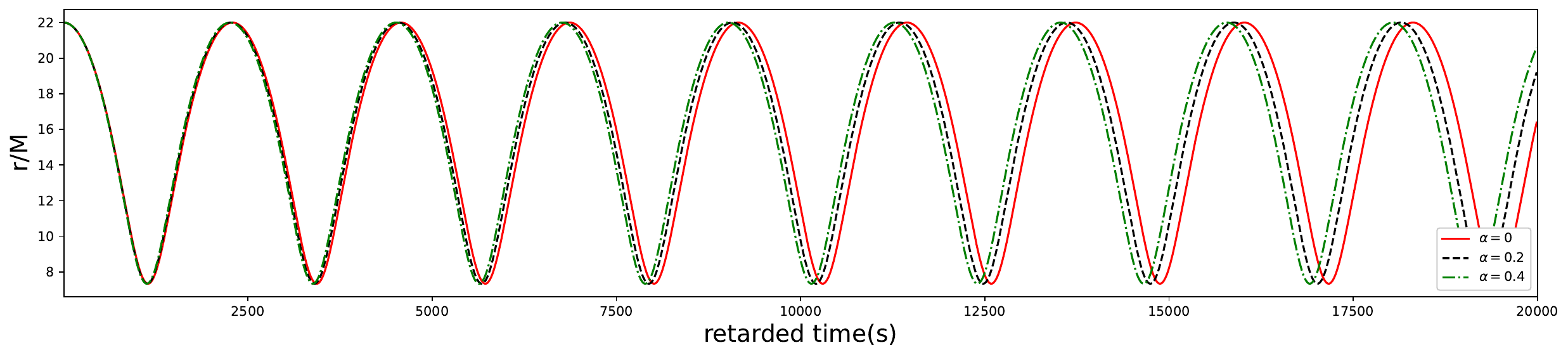}\\ \vspace{0.0cm}
\includegraphics[scale=0.35]{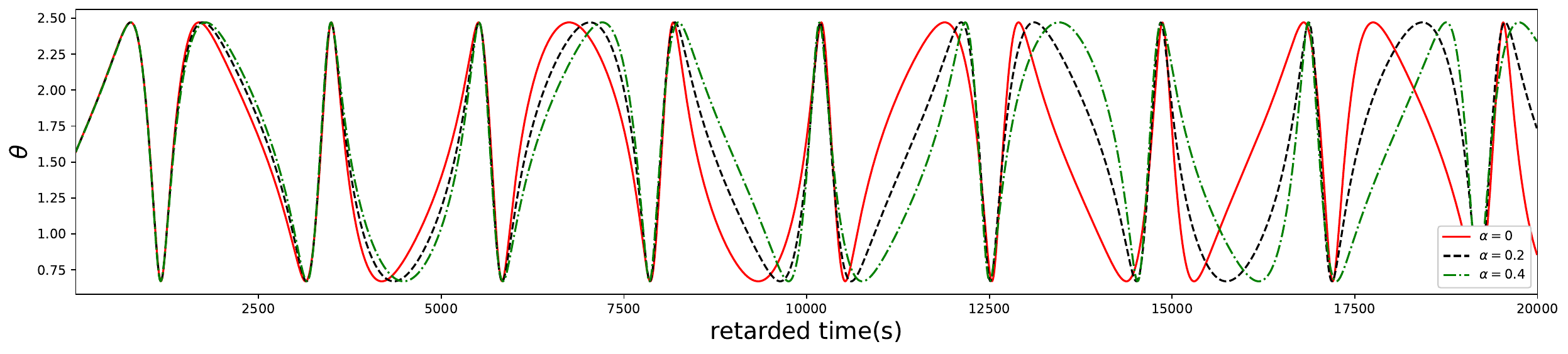}\\ \vspace{0.0cm}
\includegraphics[scale=0.35]{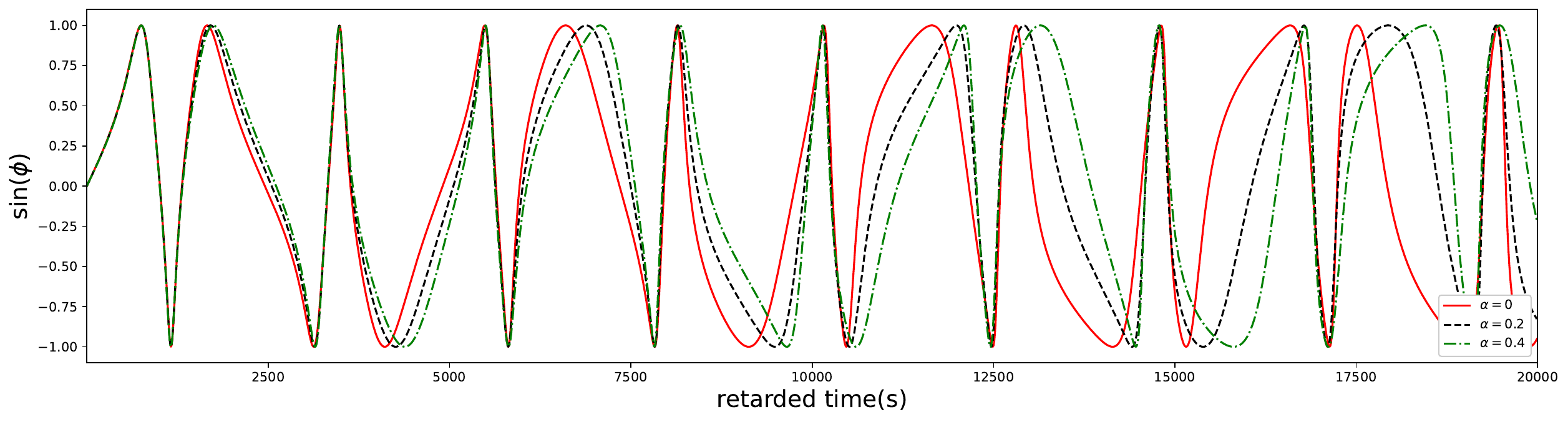}\\ \vspace{0.0cm}
\caption{\label{InclinedOrbitsRTPhi} (Color online) Time series of motion in the $r$, $\theta$ and $\phi$ directions of the Kerr-MOG orbit with various parameters $\alpha$ for the first 20000 s in the case of $(a,M,e,p,\iota)=(0.25,1\times 10^{6} M_\odot,0.5,11.0,0.9)$. 
The red, black, and green lines in each panel correspond to increasing modified parameters, i.e., $\alpha=0$, $0.2$, and $0.4$, respectively.}
\end{figure}

\begin{figure}[ht]
\includegraphics[scale=0.35]{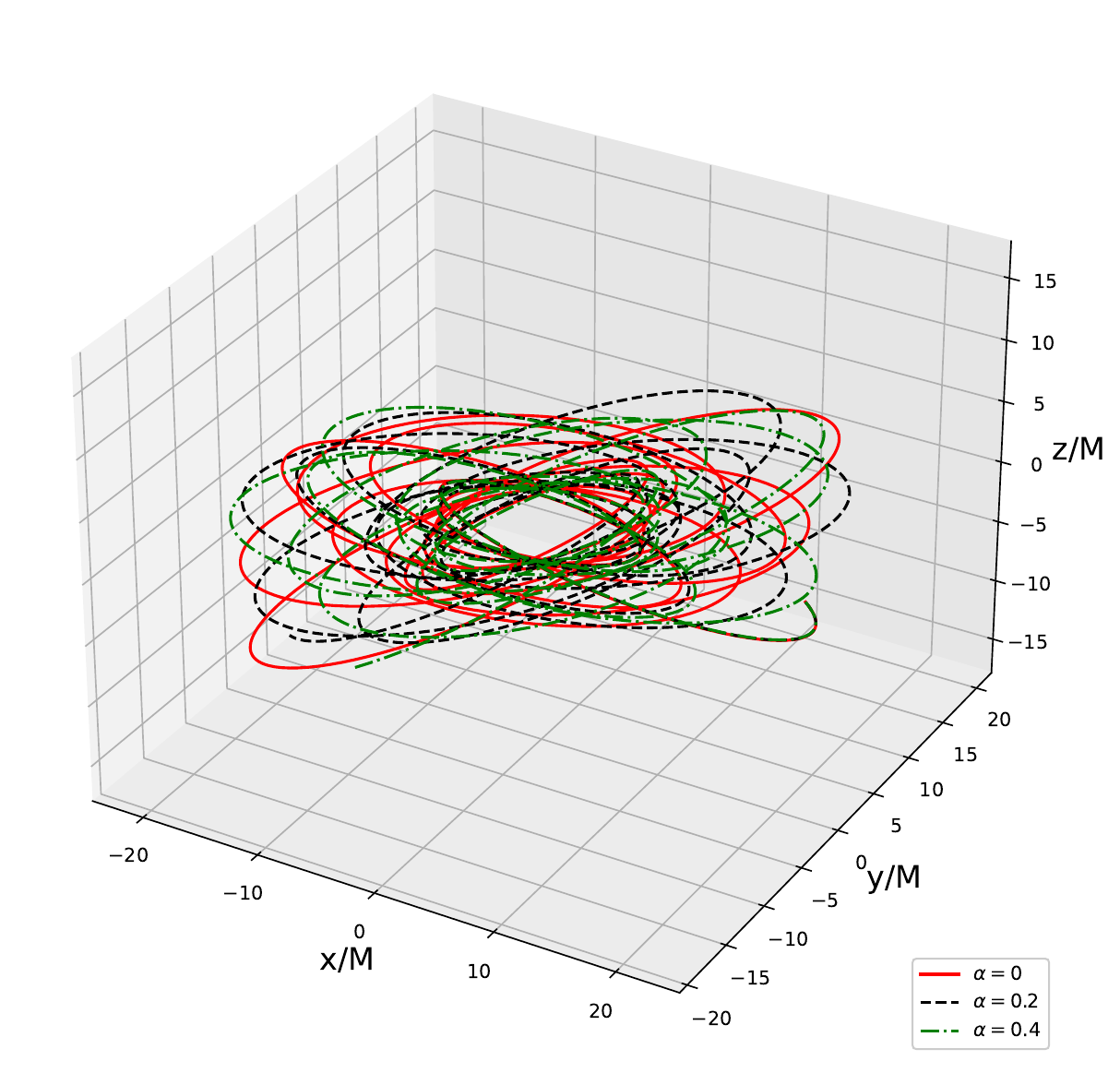}\hspace{0.2cm}%
\includegraphics[scale=0.35]{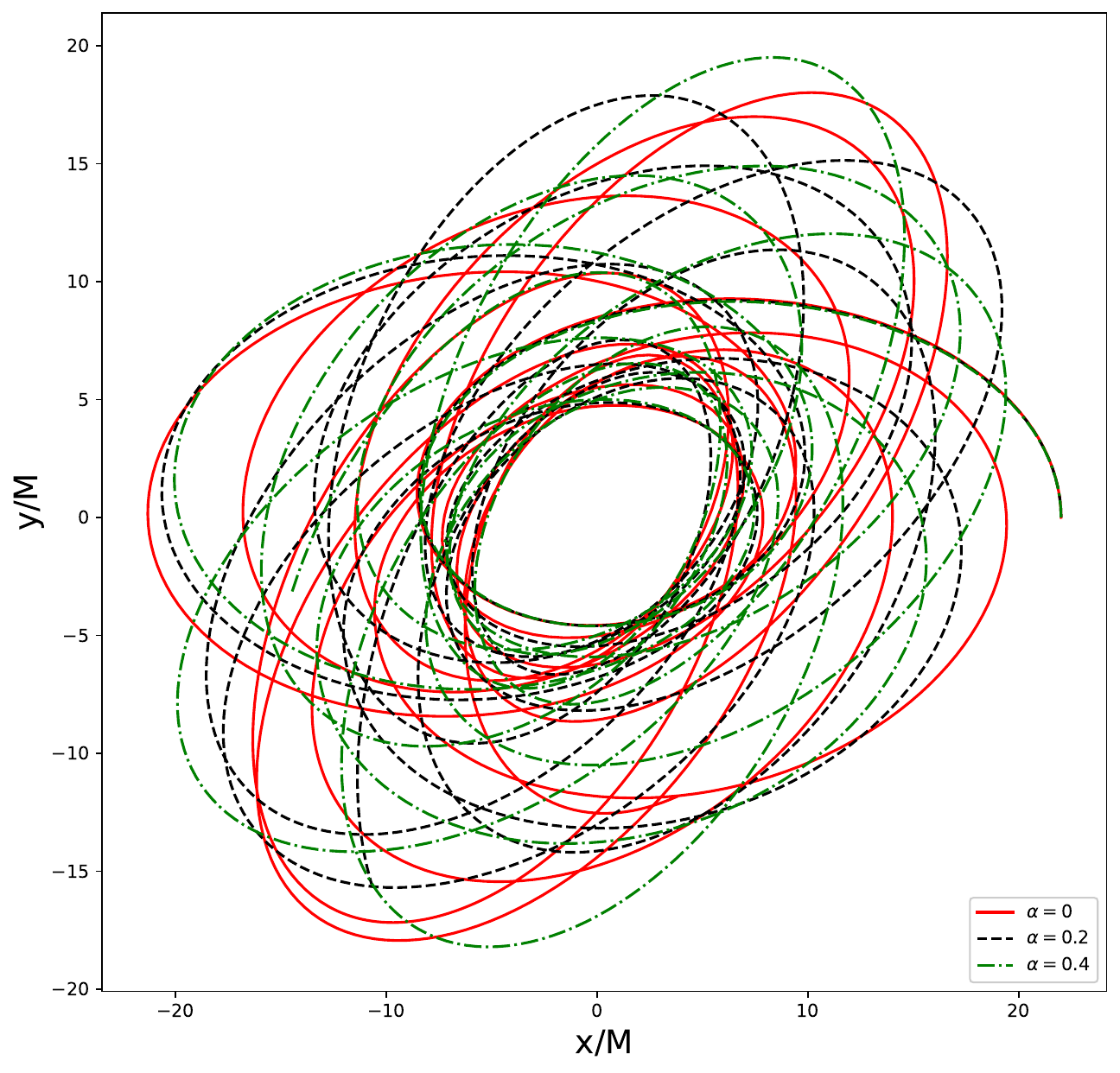}\hspace{0.2cm}%
    \includegraphics[scale=0.35]{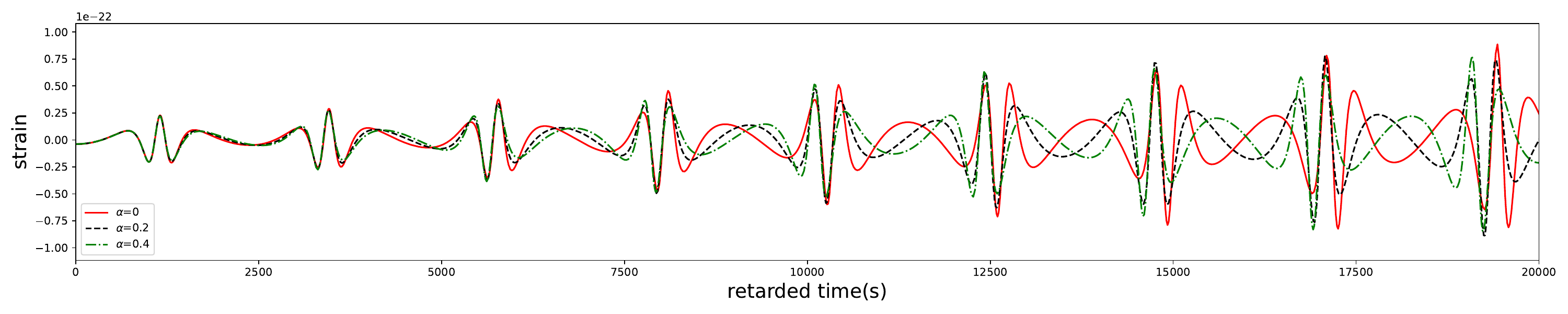}\\ \vspace{0.0cm}
\caption{\label{InclinedOrbitsGW} (Color online) Orbits and EMRI waveforms of the Kerr-MOG orbit with various parameters $\alpha$ in the case of $(a, M,e,p,\iota)=(0.25,1\times 10^{6} M_\odot,0.5,11.0,0.9)$ for the first 20000 s. The top-left panel is for the trajectories of the particle, the top-right panel is for the projection onto the $x$-$y$ plane of the trajectories, and the bottom panel is for the component $h_I$ of the EMRI waveforms. 
The red, black, and green lines in each panel correspond to increasing modified parameters, i.e., $\alpha=0$, $0.2$, and $0.4$, respectively.}
\end{figure}

The above results are valid only for equatorial geodesic orbits. We should also consider more general inclined geodesics.
To analyze inclined geodesic orbits, it is essential to account for the motion in the polar direction in addition to the radial and azimuthal directions. 
In this case, the initial values of $(E,L_z,Q)$ are obtained by solving the following equations
\begin{equation}
R(E,L_z,Q,r=r_a)=0,~~~~R(E,L_z,Q,r=r_p)=0,~~~~Q=Q(E,L_z,\theta=\theta_{min}).
\end{equation}
For the same initial positions and parameters $e$ and $p$, in Fig.~\ref{InclinedOrbitsRTPhi}, we display the time series of motion in the $r$, $\theta$, and $\phi$ directions for a Kerr-MOG orbit, exploring the effects of varying the modified parameter $\alpha$ over the first 20000 seconds.
Analogous to the motions in the $r$ and $\phi$ directions, $\alpha$ significantly influences the phase of the $\theta$ motion while having almost no effect on its amplitude.
The increase of $\alpha$ leads to a decrease of the period of motion in the $r$ direction but an increase of the periods in the $\theta$ and $\phi$ directions.
This behavior aligns with the findings shown in Fig.~\ref{fig:alpha_3D_freq}, i.e., as $\alpha$ increases, the radial frequency $\omega_{r}$ increases,  while  the azimuthal frequency $\omega_{\theta}$ and the polar frequency $\omega_{\phi}$ decrease.
Now, we are ready to generate EMRI waveforms using the kludge method. In Fig.~\ref{InclinedOrbitsGW}, we present the inclined trajectories and corresponding component $h_I$ of the EMRI waveforms with various $\alpha$ over the initial 20000 seconds.
It is found that, as the retarded time increases, the trajectories and waveforms with different $\alpha$ exhibit distinct differences.

\begin{figure}[ht]
\includegraphics[scale=0.54]{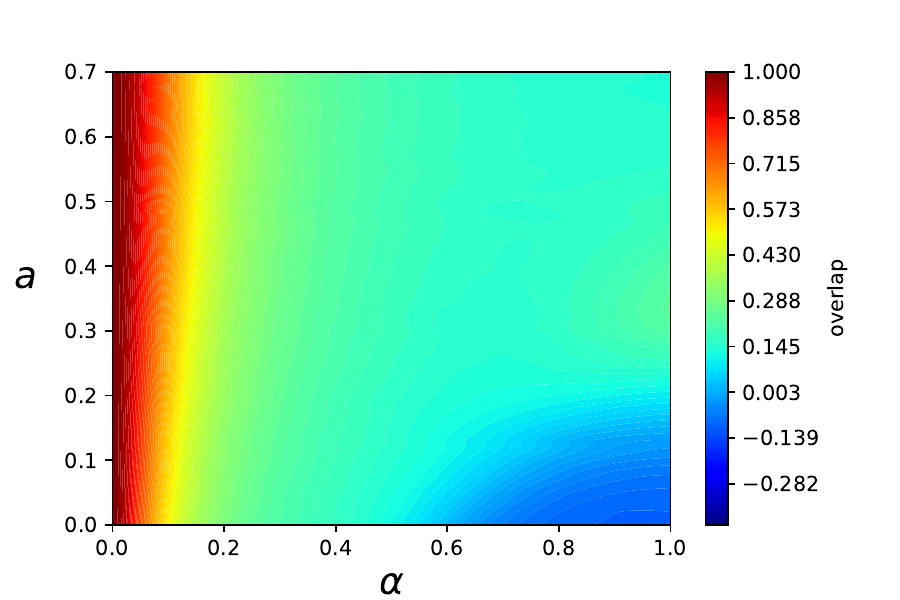}\vspace{0.0cm}
\includegraphics[scale=0.54]{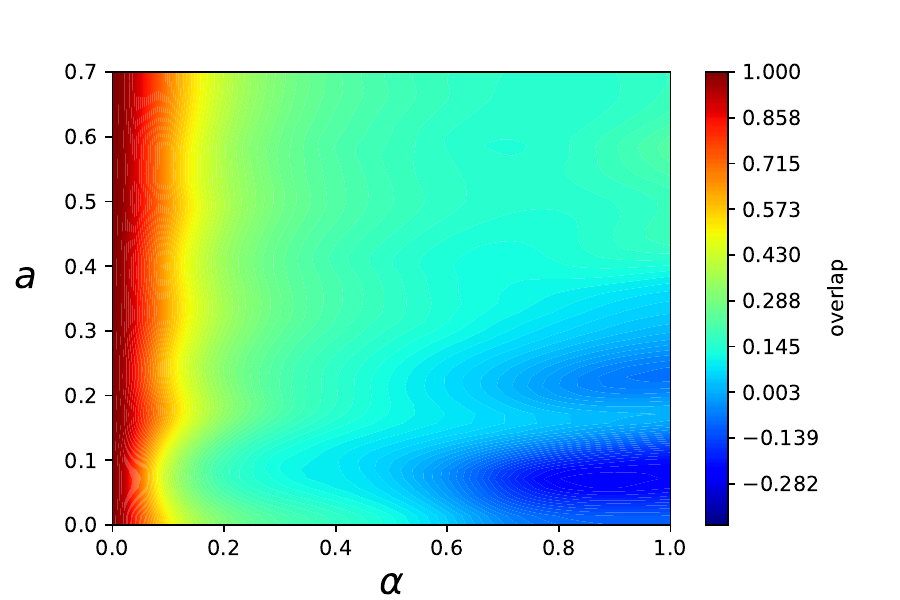} \\ \hspace{0.0cm}
\caption{\label{OverlapEquatorialInclined} (Color online) Overlap between the Kerr-MOG waveform and Kerr waveform for the equatorial geodesic orbits $(M,e,p,\iota)=(1\times 10^{6} M_\odot,0.5,11.0,0)$ (left) and inclined geodesic orbits $(M,e,p,\iota)=(1\times 10^{6} M_\odot,0.5,11.0,0.9)$ (right). The grid size is $50\times50$.}
\end{figure}


To more directly assess the impact of the modified parameter $\alpha$ on EMRIs, Fig.~\ref{OverlapEquatorialInclined} presents the overlap for the first $10^5$ s between the Kerr-MOG and Kerr waveforms for both equatorial geodesic orbits and inclined geodesic orbits in the parameter space  $(a, \alpha)$.
We observe that the overlap diminishes as the modified parameter $\alpha$ increases or the spin parameter $a$ decreases. 
This trend suggests that including higher modified parameters facilitates the differentiation of GW waveforms between a Kerr-MOG orbit and a Kerr orbit under identical initial parameters.
It should be noted that the inclination angle $\iota$ significantly influences the overlap distribution compared with equatorial geodesic orbits, as illustrated in the right panel of Fig.~\ref{OverlapEquatorialInclined}. 
This difference underscores that the combined effects of the modified parameter $\alpha$, the spin parameter $a$, and the inclination angle $\iota$ contribute to a richer physical understanding of the orbits and the corresponding EMRI waveforms. 

\subsection{Confusion problem}

When attempting to identify  EMRI signals, we encounter a significant challenge known as the confusion problem~\cite{glampedakis2006}. This issue arises from the possibility of close alignment between the gravitational waveforms of Kerr and Kerr-MOG spacetimes. Specifically, this alignment can occur when a Kerr black hole and a Kerr-MOG black hole share the same orbital frequency, leading to significant overlaps in their waveforms.
Namely, the overlaps greater than 0.97 make it difficult to distinguish between Kerr and Kerr-MOG signals when they are characterized by fully different parameters. 
It is convenient to study the confusion problem by evaluating the overlap between two waveforms at the same orbital frequencies, thereby circumventing the need to check the entire parameter space thoroughly. In this subsection, we will discuss the details of the confusion problem, 
focusing on EMRIs originating from both equatorial and inclined orbits.

\subsubsection{Confusion problem for equatorial geodesics}
From Eqs.~(\ref{EqFreqr})-(\ref{EqFreqphi}), it is evident that the frequencies are functions of the black hole parameters $(\alpha, a, M, p, e, \iota)$. For equatorial orbits, where $\iota = 0$, only $\omega_r$ and $\omega_\phi$ need to be considered. To investigate whether there is any confusion between Kerr black holes and Kerr-MOG black holes, the characteristic frequencies must be equivalent in both spacetimes. In our calculations, we will compare the frequency relationships for $\alpha=0$ and $\alpha=\alpha_0$. Therefore, only four independent parameters remain: $(a, M, e, p)$. By fixing any two of these parameters and varying the other two, we aim to achieve $\omega^{\text{MOG}} = \omega^{\text{Kerr}}$. For example, by varying $(e,p)$, we can set
\begin{eqnarray}
	\omega_{r}^{MOG}\left(\alpha_0,a_0,M_0,e_0+\delta e, p_0+\delta p,  \iota=0  \right)=\omega_{r}^{Kerr}(\alpha=0, a_0,M_0,e_0,p_0, \iota=0 ),\label{eq:omega_1}\\
	\omega_{\phi}^{MOG}\left(\alpha_0,a_0,M_0, e_0+\delta e,p_0+\delta p,  \iota=0  \right)=
 \omega_{\phi}^{Kerr}\left(\alpha=0,a_0, M_0,e_0,p_0,  \iota=0  \right).\label{eq:omega_2}
\end{eqnarray}
Alternatively, we can select $(a,M)$ as variables
\begin{eqnarray}
	\omega_{r}^{MOG}\left(\alpha_0,a_0+\delta a_0,M_0+\delta M,e_0, p_0,  \iota=0    \right)=\omega_{r}^{Kerr}(\alpha=0,a_0,M_0,e_0, p_0,  \iota=0 ),\\
	\omega_{\phi}^{MOG}\left(\alpha_0,a_0+\delta a_0,M_0+\delta M, e_0,p_0,  \iota=0    \right)=\omega_{\phi}^{Kerr}(\alpha=0,a_0,M_0,e_0,p_0,  \iota=0  ).
\end{eqnarray}

\begin{figure}[ht]
\includegraphics[scale=0.47]{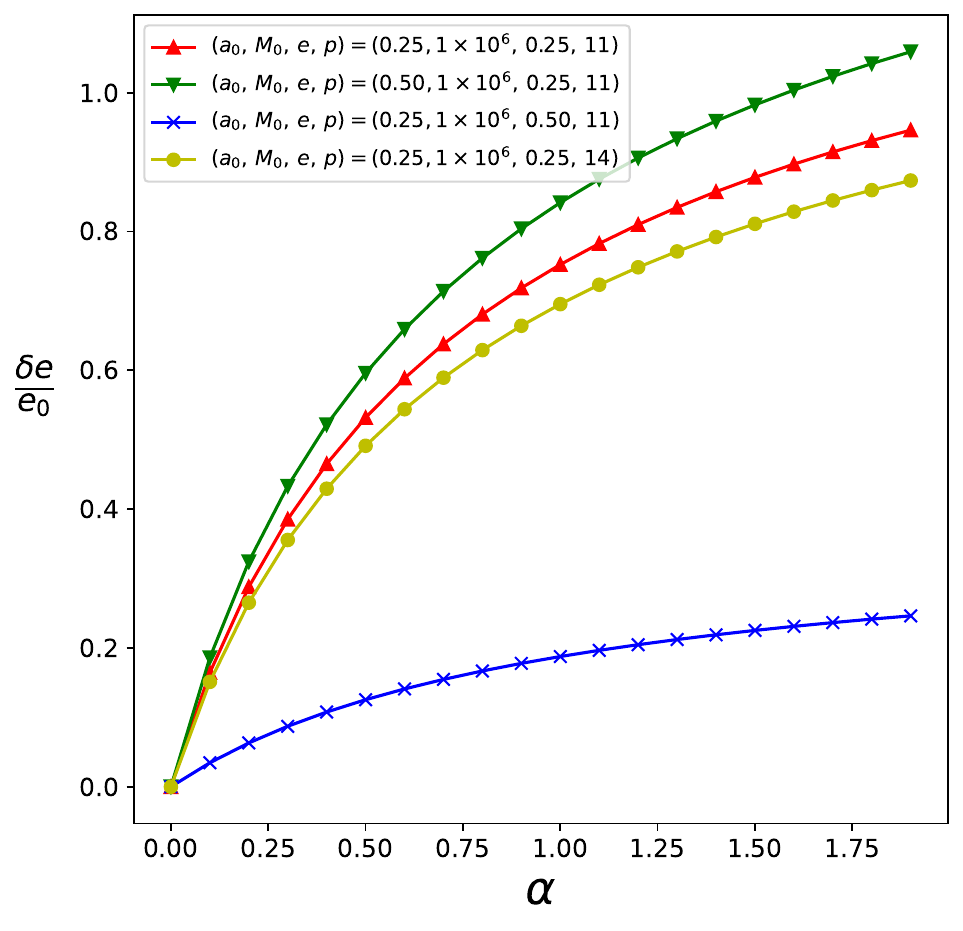}\hspace{0.25cm}%
\includegraphics[scale=0.47]{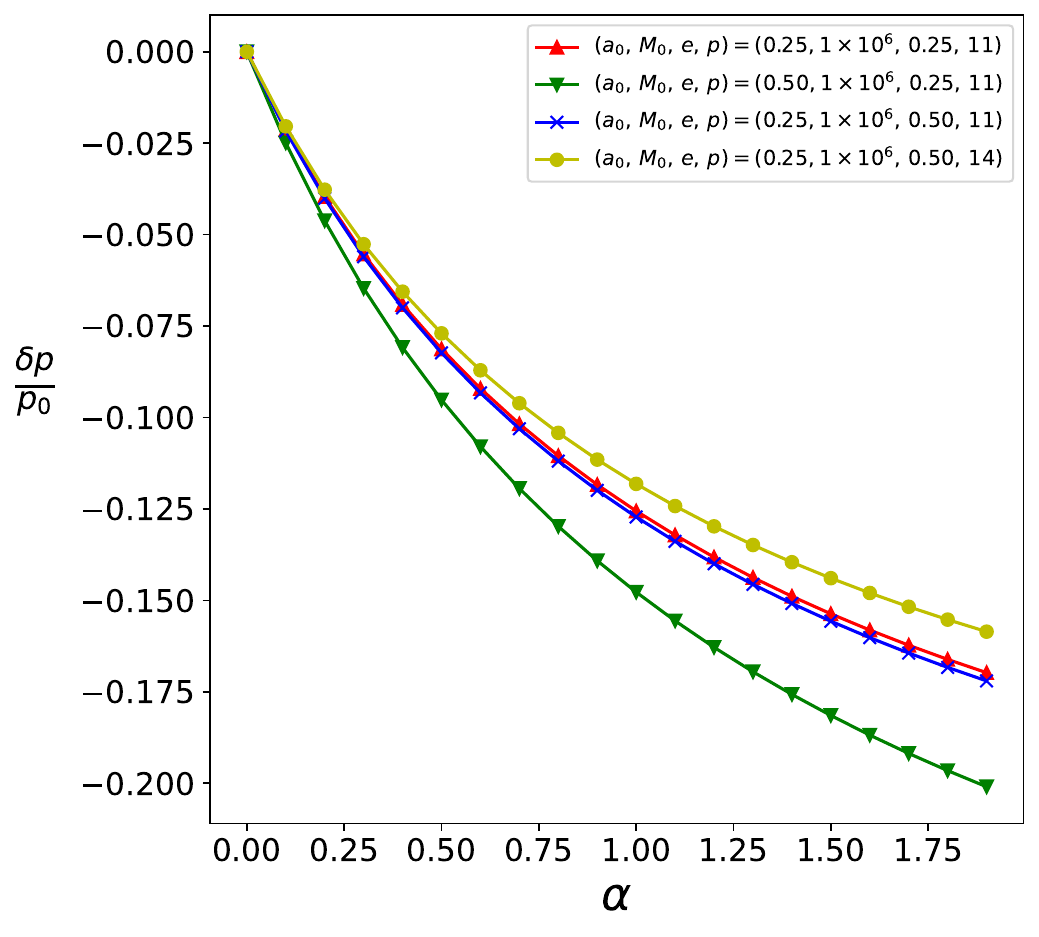}\hspace{0.25cm}%
\caption{\label{fig:e_alpha} (Color online) Relation between relative varied orbit parameters when equating orbital frequencies, i.e., the orbits of $(0, a_0, M_0,e,p)$ and $(\alpha_0, a_0, M_0, e+\delta e, p + \delta p)$ have the same orbital frequencies and here shows ${\delta a}/{a_0}$-$\alpha$ (left) and ${\delta p}/{p_0}$-$\alpha$ (right) relations. }
\end{figure}

Firstly, we focus on the confusion problem in parameter space $(e, p)$. Eqs.~(\ref{eq:omega_1}) and (\ref{eq:omega_2}) belong to an equation set with two unknowns. 
Given any ($\alpha \neq 0$, $a, M, e, p$), finding a set of $\delta p$ and $\delta e$ to satisfy the equations is always possible. 
If other parameters are fixed and only $\alpha$ is varied, then $\delta e$ and $\delta p$ will also change.
In Fig.~\ref{fig:e_alpha}, we illustrate the relationship between the relative variations of orbital parameters and $\alpha$ when matching the orbital frequency of Kerr and Kerr-MOG black holes.  
To facilitate our analysis, we define a set of parameters for a  reference line with 
$(a_0,M_0,e_0,p_0 )=(0.25,1\times 10^5M_\odot,0.25,11)$.
Other lines in the figure represent variations in $a_0$, $e_0$, or $p_0$ independently.
The left panel depicts the relationship of ${\delta e}/{e_0}$ with $\alpha$, while the right panel presents the relationship of ${\delta p}/{p_0}$ with $\alpha$.
The obtained results reveal that  ${\delta e}/{e_0}$  increases rapidly and ${\delta p}/{p_0}$ decreases rapidly as $\alpha$ increases, indicating that variations in $\alpha$ necessitate adjustments in the orbital parameters $e$ and $p$ to preserve consistent orbital frequencies. 
The divergence of these lines underscores the sensitivity of waveform confusion to the parameters $(a, e,p)$.

\begin{figure}[ht]
\includegraphics[scale=0.47]{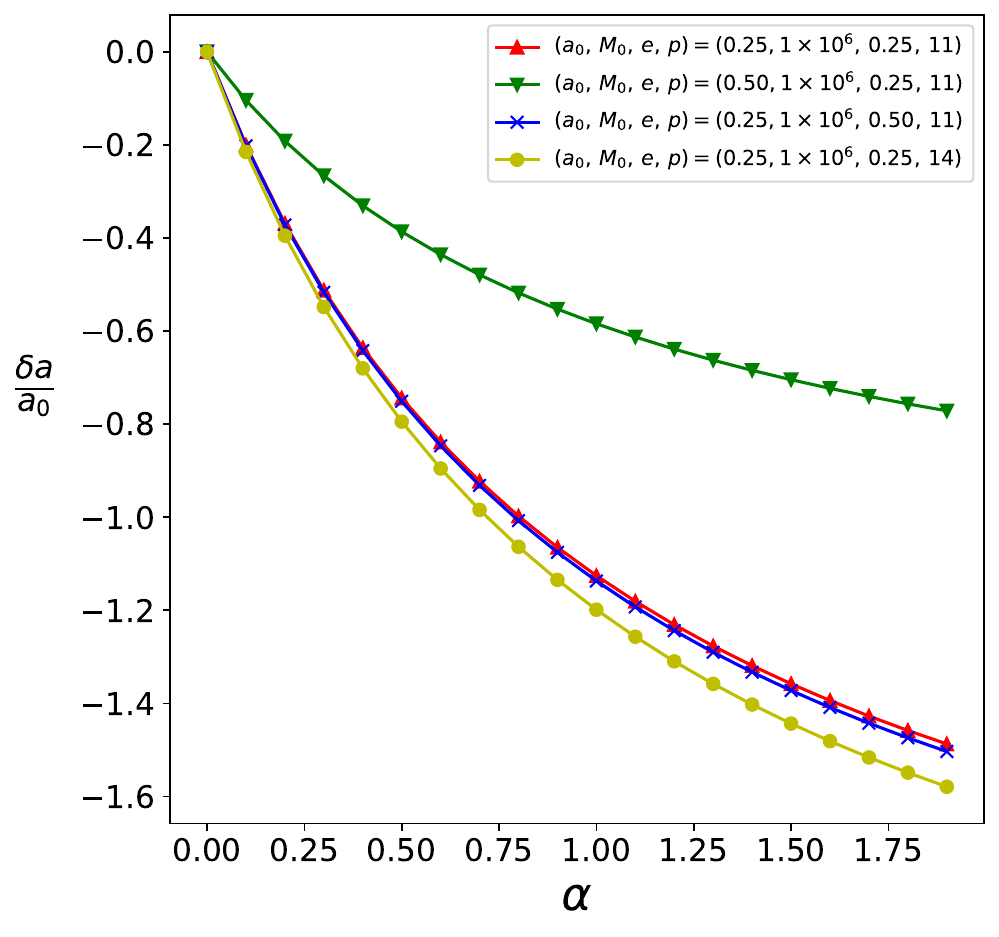}\hspace{0.25cm}%
\includegraphics[scale=0.47]{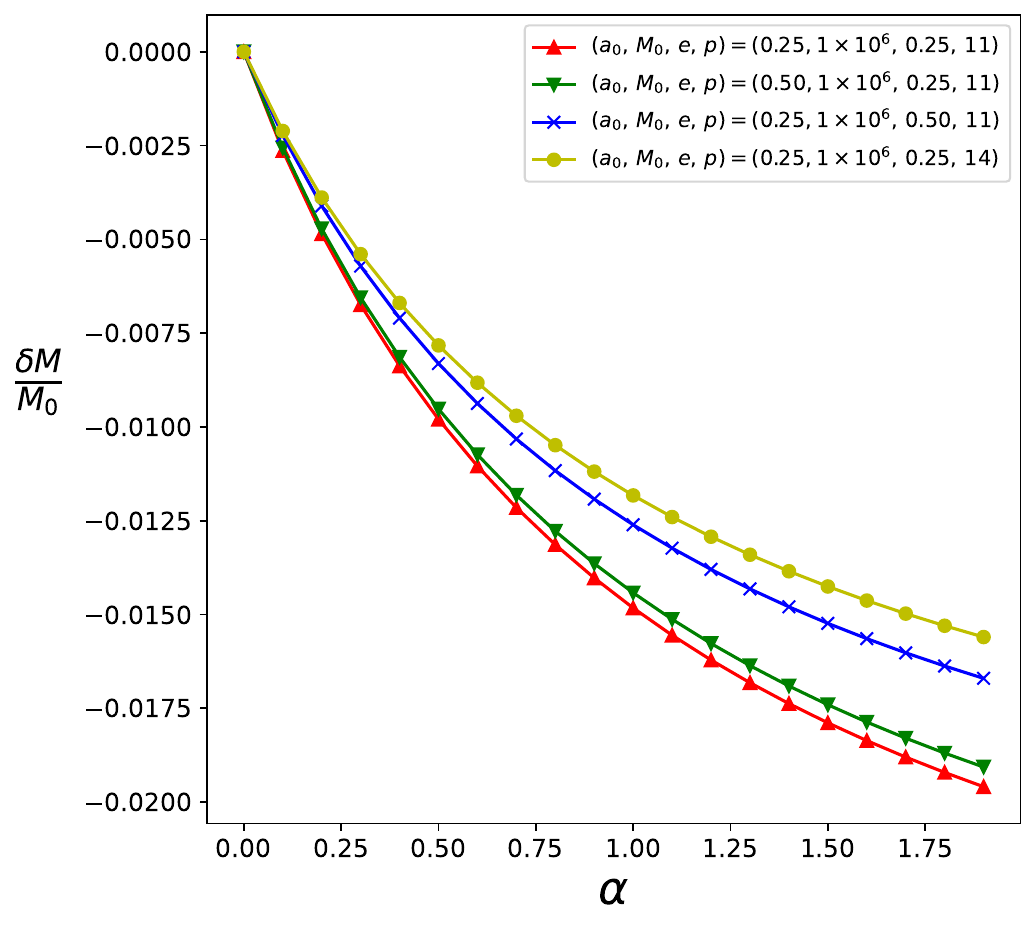}\hspace{0.25cm}%
\caption{ (Color online) Relation between relative varied black hole parameters when equating orbital frequencies, i.e., the orbits of $(0, a_0, M_0,e,p)$ and $(\alpha, a_0+ \delta a, M_0+ \delta M, e, p)$ have the same orbital frequencies and here shows $\delta a/a_0$-$\alpha$ (left) and $\delta M/M_0$-$\alpha$ (right) relations. }
\label{fig:a_alpha}
\end{figure}

Secondly, the variations in the angular momentum $a$ and the mass $M$ of the black hole can similarly induce the confusion of waveforms when attempting to match the Kerr-MOG waveforms to those of Kerr. Same with the first case, the baseline reference parameters are defined as $(a_0, M_0, e_0, p_0)=(0.25,1\times 10^5,0.25,11)$. 
In Fig. \ref{fig:a_alpha}, we display the relationship between the relative variations in black hole parameters when matching the orbital frequencies.
The left panel represents the relations of ${\delta a}/{a_0}$ with $\alpha$, while the right panel delineates the relations of  ${\delta M}/{M}$ with $\alpha$.
The obtained results reveal that  ${\delta a}/{a_0}$ and ${\delta M}/{M_0}$ decreases rapidly as $\alpha$ increases. 
The divergence of these lines also underscores the sensitivity of waveform confusion to the parameters $(a, e,p)$.

\begin{figure}[ht]
\includegraphics[scale=0.35]{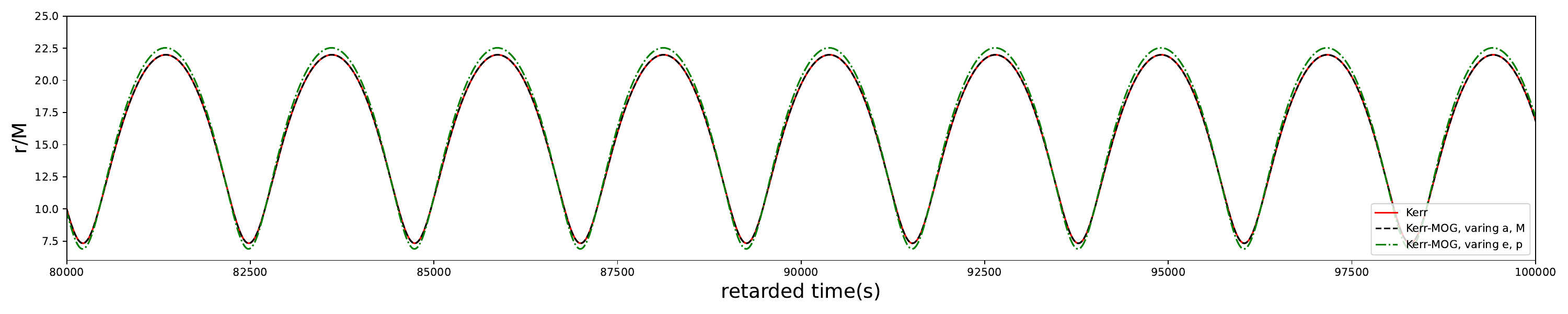}\\ \vspace{0.0cm}
\includegraphics[scale=0.35]{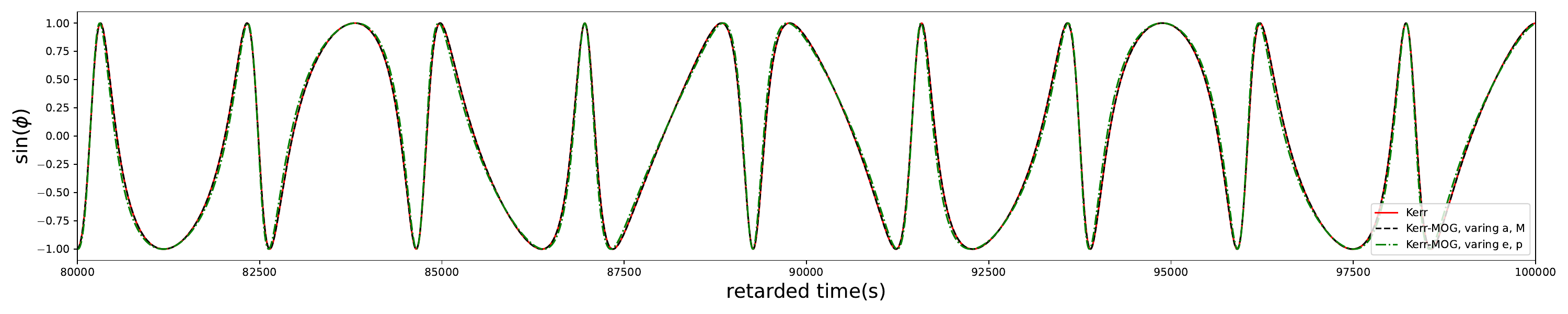}\\ \vspace{0.0cm}
\caption{ (Color online)  Time series of motion in the $r$ and $\phi$ directions of the Kerr-MOG and Kerr orbits with various parameters. The depicted lines correspond to sets of orbital and black hole parameters as follows: the red solid line represents $(\alpha, a, M, e, p, \iota) = (0, 0.25, 1 \times 10^{6} M_{\odot}, 0.5, 11.0, 0)$, the black dashed line corresponds to $(\alpha, a, M, e, p, \iota) = (0.2, 0.15699, 0.99589 \times 10^{6} M_{\odot}, 0.5, 11.0, 0)$, and the green dotted line denotes $(\alpha, a, M, e, p, \iota) = (0.2, 0.25, 1 \times 10^{6} M_{\odot}, 0.53156, 10.55899, 0)$.    }
\label{fig:equat_orbit}
\end{figure}

\begin{figure}[ht]
\includegraphics[scale=0.35]{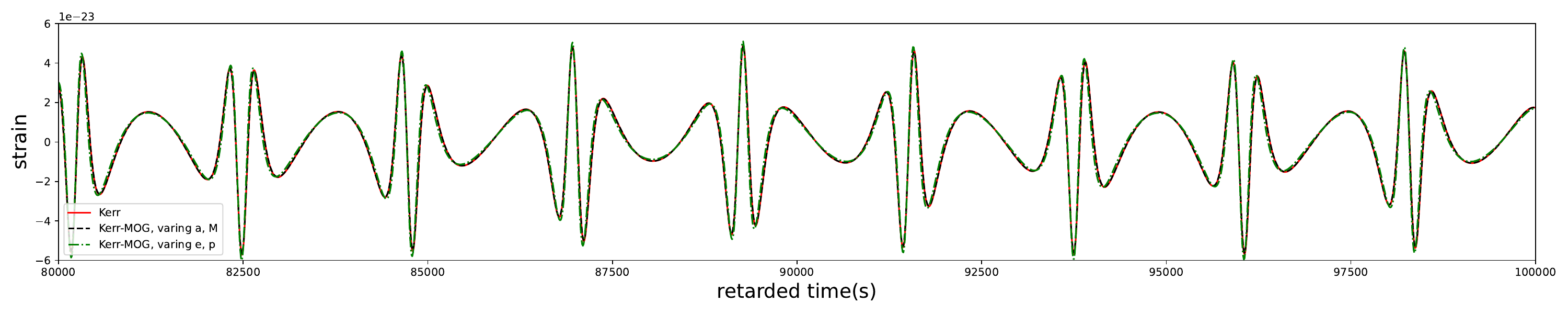}\\ \vspace{0.0cm}
\includegraphics[scale=0.35]{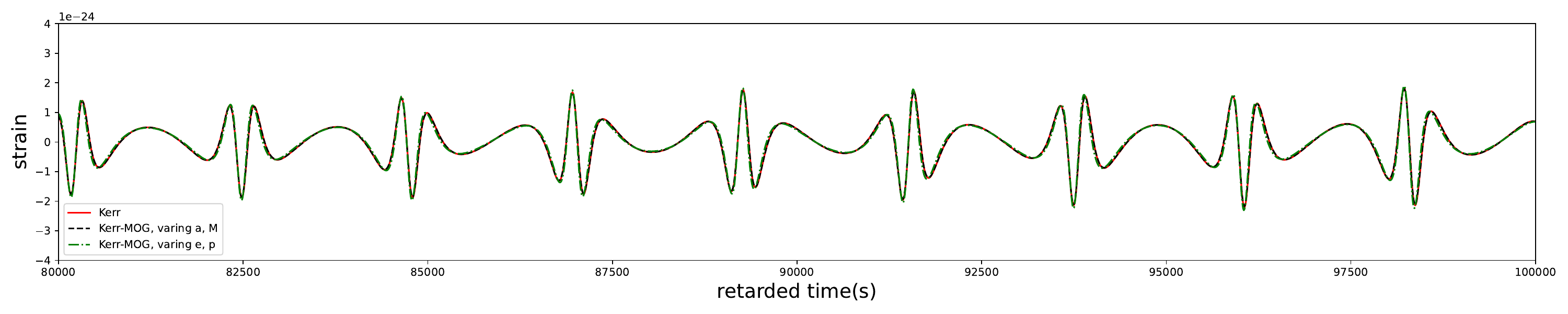}\\ \vspace{0.0cm}
\caption{(Color online) 
Waveforms for the Kerr-MOG and Kerr orbits with the same parameters as in Fig. \ref{fig:equat_orbit}, detailing the strain over the retarded time for the interval 80000 - 100000 s. The top panel illustrates the strain in channel I of the LISA data stream, while the bottom panel shows channel II's response. }
\label{fig:equat_waveform}
\end{figure}

In the previous paragraphs, we have established that the orbital frequencies of Kerr and Kerr-MOG systems are identical under specific conditions. This finding prompts a further investigation into whether this frequency equivalence translates to similarities of the geodesic orbits. 
In Fig.~\ref{fig:equat_orbit}, we present the $r$ and $\theta$ motion of Kerr-MOG black hole 
compared to the Kerr black hole. 
Although there are slight deviations at the extrema in the radial direction, the radial and azimuthal components of orbits are almost identical after 80000s.
In fact, since their orbital frequencies are identical, these orbits nearly completely overlap and remain consistent over time without diverging.
With NK method, a specific  orbit of the secondary corresponds to a definite GW waveform.
Therefore, in theory, the significant overlap in the orbits suggests that the GW waveforms will likewise exhibit a high degree of similarity.
In Fig.~\ref{fig:equat_waveform}, we further investigate the waveforms by showcasing the strain responses in both of LISA data stream channels I and II~\cite{Canizares:2012is}, using the same $(a, M,e,p)$ parameters featured in Fig.~\ref{fig:equat_orbit}. 
The nearly identical waveforms in both the top and bottom panels emphasize the difficulty in distinguishing between Kerr and Kerr-MOG black holes through waveform analysis alone.

\begin{figure}[ht]
\includegraphics[scale=0.48]{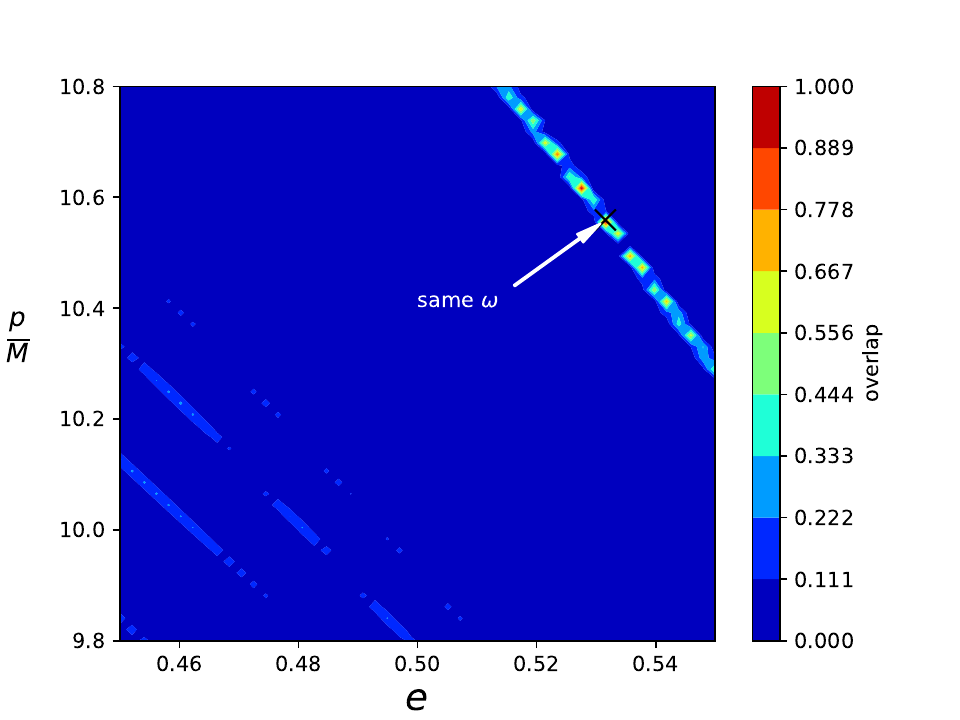}\hspace{0.25cm}%
\includegraphics[scale=0.48]{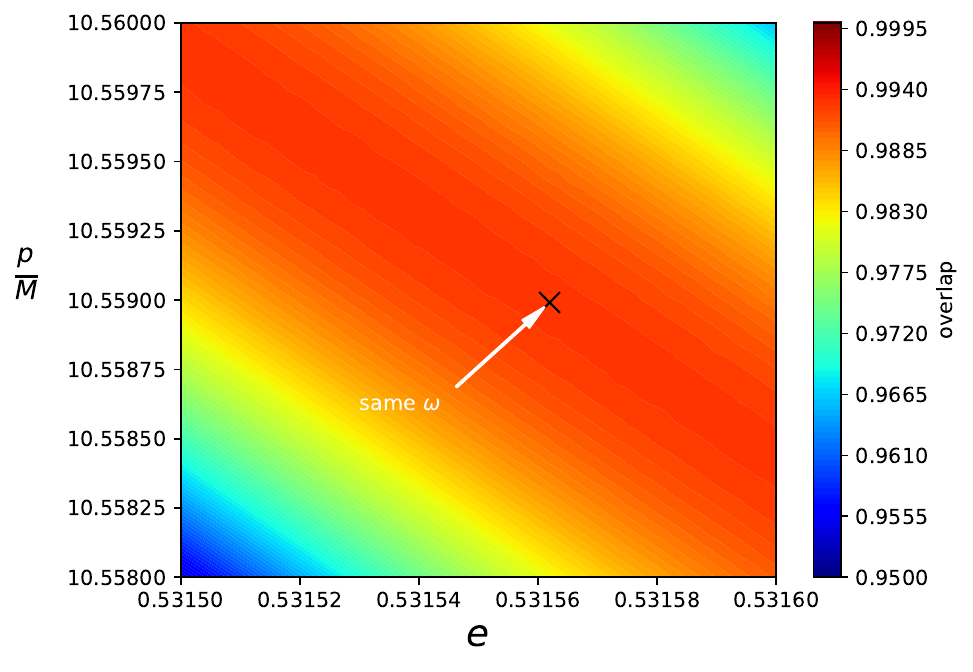}\hspace{0.25cm}%
\caption{\label{fig:e_prc } (Color online) Distribution of the overlap between waveforms defined by $(\alpha,a_0, M_0,e_0,p_0)=(0,0.25,1\times 10^{6} M_\odot,0.5,11.0)$ and  $(\alpha, a, M,e,p)=(0.2,0.25,1\times 10^{6} M_\odot,e,p)$ plane. The black cross mark is pointed by the arrow: the same $\omega_r$ and $\omega_\phi$ at ($e_{\rm{MOG}}$, $p_{\rm{MOG}}) = (0.53156,10.55900)$.  
The grid size is $50\times 50$. }
\end{figure}

Encouraged by these findings, we extend our investigation to explore the degree of overlap.
In Fig.~\ref{fig:e_prc }, we present a detailed analysis of the waveform overlap for the first $10^5$ s. The left panel offers a comprehensive overview of the overlap distribution across a spectrum of parameters for $e$ and $p/M$. 
The right panel closely examines a particular region of interest, highlighting a subset where the orbital frequencies $\omega_r$ and $\omega_{\phi}$ are identical for specific values of $e$ and $p/M$. The black cross mark, denoted as $(e_{\rm{MOG}}, p_{\rm{MOG}}) = (0.53156,10.55900)$ in the Kerr-MOG spacetime, represents the point that the overlap between the Kerr-MOG and Kerr waveforms exceeds $0.97$, rendering it exceedingly challenging to distinguish between the two waveform types.

Through the analysis of this section, we reach an important conclusion: EMRI  systems with different parameters cannot be distinguished if their orbital frequencies are the same. Taking  Eq. \eqref{eq:omega_1} and Eq. \eqref{eq:omega_2} as an example, given any ($\alpha \neq 0, a, M, e, p$), it is always possible to find a set of $\delta p$ and $\delta e$ such that the equations are satisfied. 
This means that for any given  Kerr-MOG black holes,  corresponding Kerr black holes can always be found whose EMRI waveforms are almost identical.
This indistinguishability necessitates additional analysis to resolve such confusion problems.

\subsubsection{Confusion problem for inclined geodesics}

\begin{figure}[ht]
\includegraphics[scale=0.35]{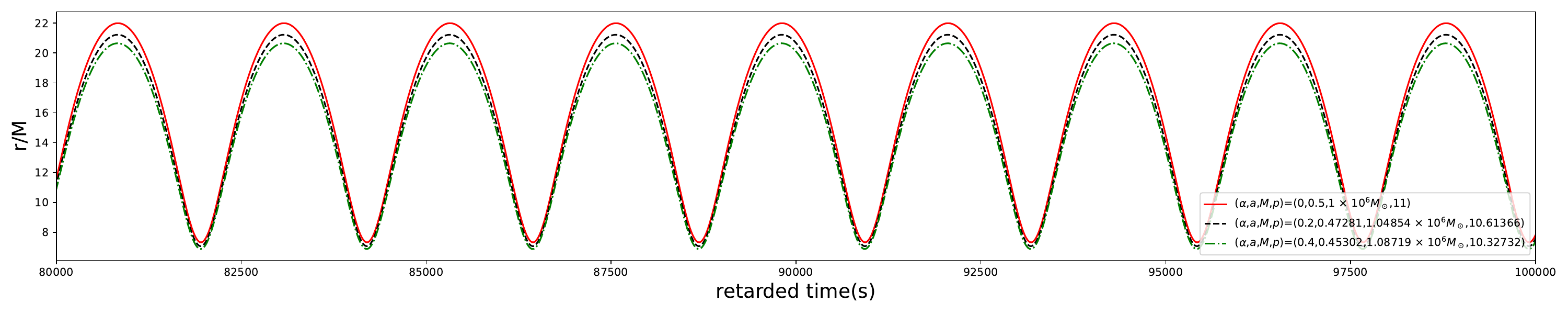}\\ \vspace{0.0cm}
\includegraphics[scale=0.35]{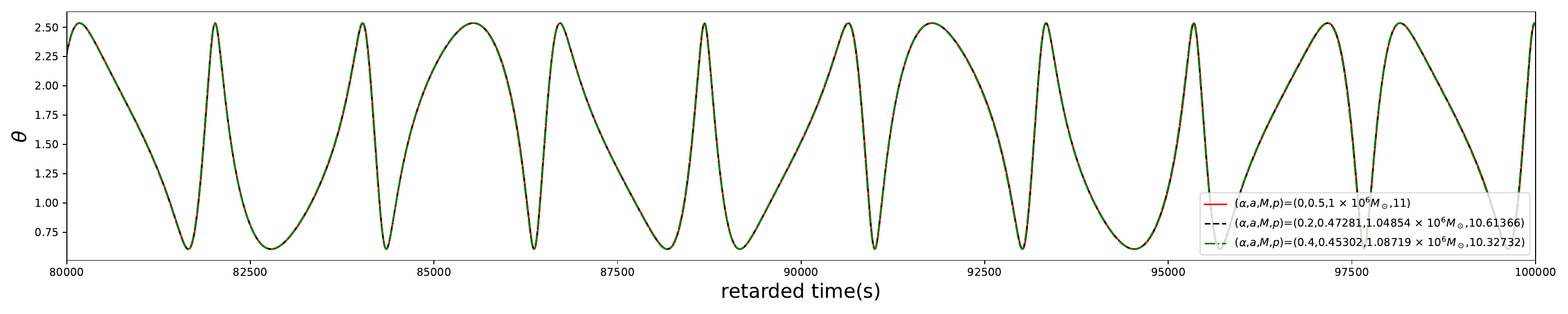}\\ \vspace{0.0cm}
\includegraphics[scale=0.35]{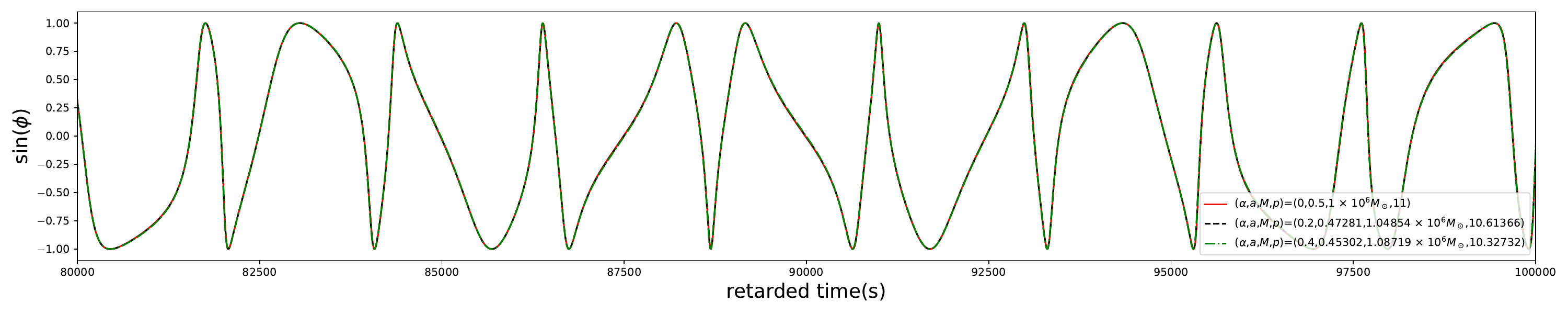}\\ \vspace{0.0cm}
\caption{ (Color online) 
Time series of motion in the $r$, $\theta$, and $\phi$ directions of the Kerr-MOG and Kerr orbits with various parameters. The depicted lines correspond to sets of orbital and black hole parameters as follows: the red solid line represents $(\alpha, a, M, e, p, \iota)=(0, 0.5, 1\times 10^{6}M_\odot, 0.5, 11, 0.965)$, the black dashed line corresponds to $(\alpha, a, M, e, p, \iota)=(0.2, 0.47281, 1.04854\times 10^{6}M_\odot, 0.5, 10.61366, 0.965)$ and the green dash-dot line denotes $(\alpha, a, M, e, p, \iota)=(0.4, 0.45302, 1.08719\times 10^{6}M_\odot, 0.5, 10.32732, 0.965)$.}
\label{fig:inclined_orbit}
\end{figure}

\begin{figure}[ht]
\includegraphics[scale=0.35]{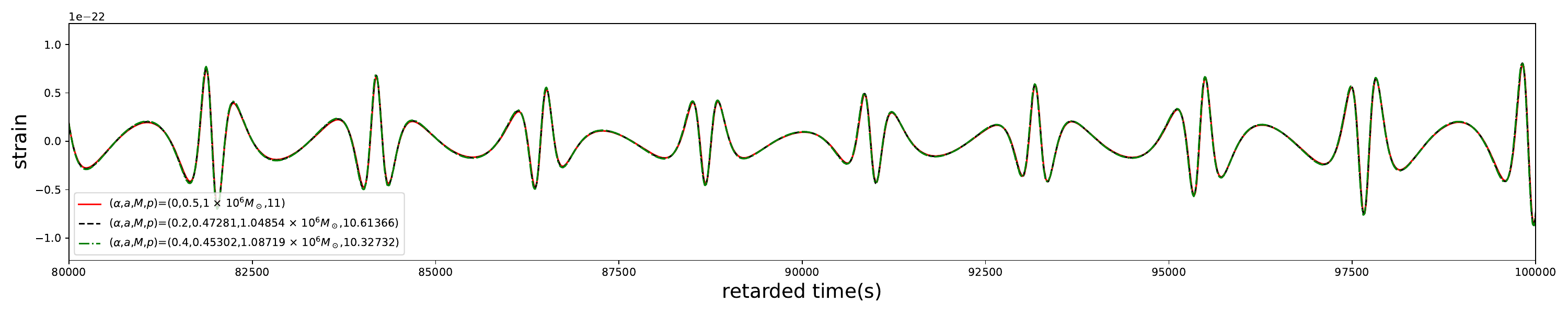}\\ \vspace{0.0cm}
\includegraphics[scale=0.35]{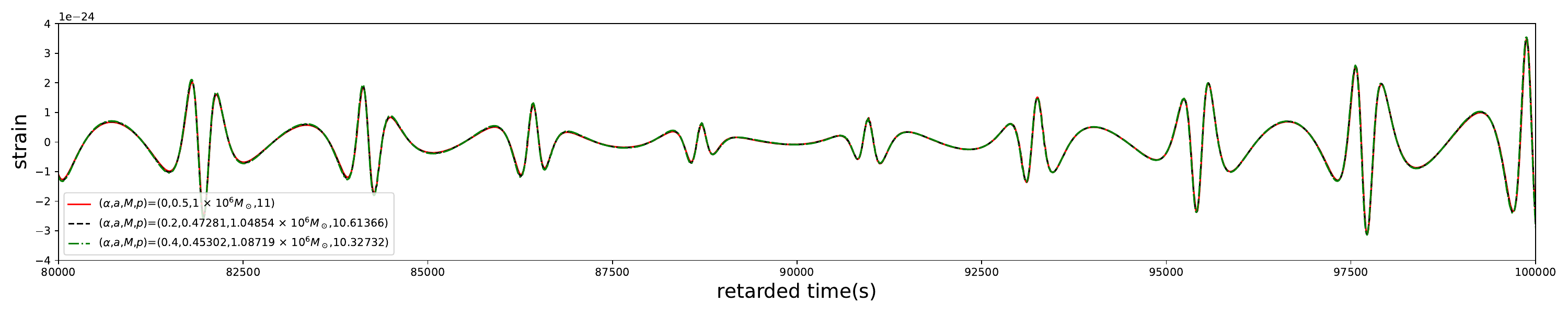}\\ \vspace{0.0cm}
\caption{ (Color online) 
Waveforms for the Kerr-MOG and Kerr orbits with the same parameters as in Fig. \ref{fig:inclined_orbit}, detailing the strain over the retarded time for the interval 80000 - 100000 s. The top panel illustrates the strain in channel I of the LISA data stream, while the bottom panel shows channel II's response.}
\label{fig:inclined_waveform}
\end{figure}

For inclined orbits, where the secondary exhibits the polar motion, it is also necessary to consider $\omega_\theta$. In this scenario, for a given $\alpha$, there are five independent variables: $(a, M, e, p, \iota)$. We can fix two of these variables and vary the remaining three  to investigate the confusion problem. Here we choose to vary $(a, M, p)$. As a matter of fact, the other choices will not qualitatively modify our results. Under these conditions, the criterion for equal frequencies is given by
\begin{eqnarray}
\omega_{r}^{MOG}\left(\alpha_0, a_0+\delta a,M_0+\delta M,e_0,p_0+\delta p,  \iota_0  \right)=\omega_{r}^{Kerr}(\alpha=0, a_0,M_0, e_0,p_0, \iota_0 ),\\
\omega_{\theta}^{MOG}
\left(\alpha_0, a_0+\delta a,M_0+\delta M,e_0,p_0+\delta p,  \iota_0   \right)
=\omega_{\theta}^{Kerr}(\alpha=0, a_0,M_0,e_0, p_0,  \iota_0  ),\\
\omega_{\phi}^{MOG}\left(\alpha_0, a_0+\delta a,M_0+\delta M,e_0,p_0+\delta p,  \iota_0  \right)=\omega_{\phi}^{Kerr}(\alpha=0, a_0,M_0,e_0, p_0, \iota_0).
\end{eqnarray}
Given any set of parameters $(\alpha_0, a_0, M_0, e_0, p_0, \iota_0)$, we can determine the corresponding values of $(\delta a, \delta M, \delta p)$. By substituting these values into the geodesic equations, we can derive the associated orbit. For example, substituting $(\alpha, a_0, M_0, e_0, p_0, \iota_0)=(0.2, 0.5, 1\times 10^{6}M_\odot, 0.5, 11, 0.965)$ into the above formula to solve for $(\delta a, \delta M, \delta \iota)$, we obtain the parameters of Kerr-MOG black hole
$(\alpha, a, M, e, p, \iota)=(0.2, 0.47281, 1.04854\times 10^{6}M_\odot, 0.5, 10.61366, 0.965)$ and $(\alpha, a, M, e, p, \iota)=(0.4, 0.45302, 1.08719\times 10^{6}M_\odot, 0.5, 10.32732, 0.965)$. In Fig.~\ref{fig:inclined_orbit}, we depict the  $r$, $\theta$, and $\phi$ motion when the orbital frequencies match.
As illustrated in the figure, increasing $\alpha$ reduces the  amplitude of motion in the $r$ direction without affecting the motion in the $\theta$ and $\phi$ directions.
The related GW waveforms  are presented in Fig.~\ref{fig:inclined_waveform}.
The figure shows that the more general inclined orbits still have the confusion problem, meaning that we cannot distinguish whether an GW waveform comes from a Kerr-MOG black hole or a Kerr black hole.

\begin{figure}[ht]
\includegraphics[scale=0.4]{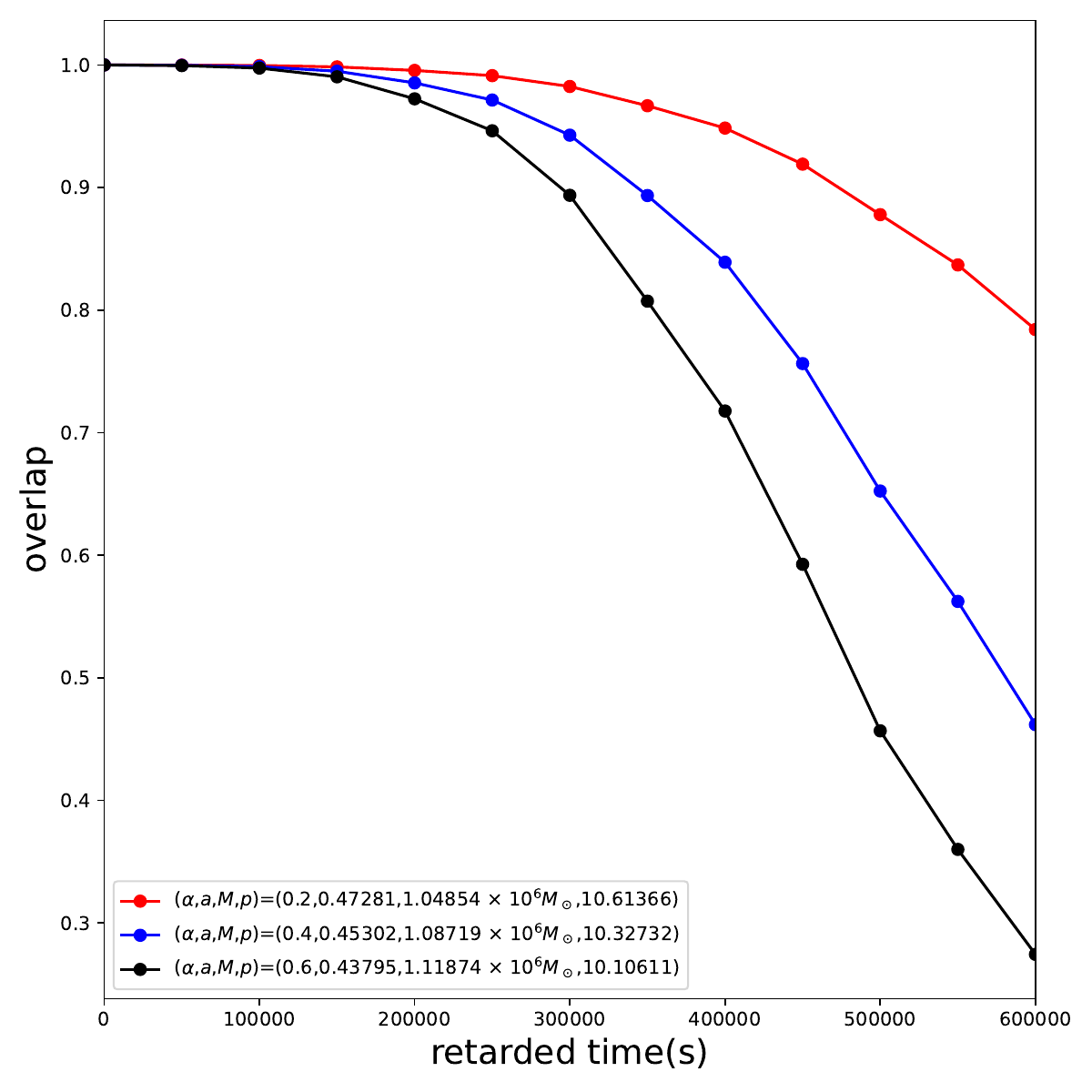}\hspace{0.2cm}%
\caption{ (Color online)  Overlap as a function of retarded time. The parameters of Kerr black hole are $(\alpha, a_0, M_0, e_0, p_0, \iota_0)=(0, 0.5, 1\times 10^{6}M_\odot, 0.5, 11, 0.965)$, corresponding parameters of Kerr-MOG black hole are denoted in the label.}
\label{fig:overlap_time}
\end{figure}

To summarize our findings, our analysis reveals that the waveforms for specific parameter sets exhibit a high degree of congruence, with the overlap between the Kerr-MOG and Kerr waveforms exceeding 0.97. 
This severely hinders our ability to detect the underlying gravitational theory by analyzing GW waveforms.

However, it is crucial to emphasize that a prominent advantage of EMRI GW detection lies in the accumulative effect over an extended period, which can amplify even theoretically small deviations. 
Therefore, a reasonable attempt to address the confusion problem is to prolong the duration of the GW observation. 
For a longer period of observation, the radiation reaction plays a significant role in simulating the EMRI waveforms.
After taking into account the radiation reaction, as an example illustrated in Fig.~\ref{fig:overlap_time}, we have calculated the relationship between the GW overlap and the evolution period for parameter sets initially exhibiting the confusion problem, with $\alpha=0.2,\ 0.4$ and $0.6$, respectively.
As illustrated, considering the effects of radiation reaction reveals a progressive decline in the overlap over time. 
The reduction in the overlap becomes more pronounced with the increase in retarded time, suggesting a viable strategy for distinguishing between the waveforms.
Moreover, the figure also indicates that a more significant $\alpha$ results in a more rapid decay of overlap. 
For instance, with $\alpha=0.2$, the overlap diminishes to approximately 0.78 after $6\times 10^5$ seconds, whereas with $\alpha=0.6$, it drops to around 0.27 in the same time interval.
The parameter $\alpha$, intrinsic to the Kerr-MOG model, plays a pivotal role: the greater its value, the more pronounced the separation between the waveforms. 
This implies that for Kerr-MOG theories with smaller deviations from the Kerr result, the EMRI GW detection will face more significant challenges in constraining and detecting the modified parameter $\alpha$.

\section{Concluding Remarks\label{chapt:5}}

In this study, we conduct a comprehensive analysis of EMRI GWs within the framework of Kerr-MOG spacetimes.
We obtained EMRI waveforms using the NK method and found that, for the same parameters $(a, M,e,p)$, the EMRI waveforms of Kerr-MOG black holes are rather different from those of Kerr black holes.
This indicates that the parameter $\alpha$ significantly affects the geodesic orbits and EMRI waveforms, allowing for a clear distinction between Kerr-MOG and Kerr waveforms.
However, one encounters the following confusion problem: for any given Kerr-MOG EMRI waveform, there is always a corresponding, essentially indistinguishable waveform of a Kerr black hole associated with different underlying parameters.
This similarity poses a significant challenge when using the waveforms to distinguish between GR and MOG.

To address the waveform confusion problem, we further scrutinized its causes.
The confusion is found to arise when the radial, angular, and azimuthal frequencies of the geodesic orbits are almost identical.
This results in Kerr and Kerr-MOG black holes with different parameters $(a, M,e,p)$ but identical geodesic orbits, leading to indistinguishable EMRI waveforms.
To resolve the confusion problem, we incorporated the effect of radiation reaction and observed a substantial reduction in the waveform overlap over time, significantly improving the ability to distinguish between different waveforms over extended periods.

We argue that the findings of the present study enhance our understanding of the observational implications of MOG parameters and the impact of the radiation reaction on distinguishing different alternative theories of gravity through measured gravitational waveforms.
More studies are warranted to dive deeper into the line of research.
In particular, it is potentially promising to perform relevant analyses using Bayesian and machine learning-based algorithms.
We plan to continue investigating this topic in future studies.

\begin{acknowledgments}

We thank Profs. Wen-Biao Han, Bin Wang, and Rui-Hong Yue for their helpful discussions and comments. 
This work is supported by the National Key Research and Development Program of China (Grant No. 2020YFC2201400), National Natural Science Foundation of China (Grant Nos. 12275079 and 12035005), innovative research group of Hunan Province (Grant No. 2024JJ1006) and Fundamental Research Program of Shanxi Province (Grant Nos. 202303021212296 and 202203021222308). 
This work is also supported by the financial support from Brazilian agencies Funda\c{c}\~ao de Amparo \`a Pesquisa do Estado de S\~ao Paulo (FAPESP), Funda\c{c}\~ao de Amparo \`a Pesquisa do Estado do Rio de Janeiro (FAPERJ), Conselho Nacional de Desenvolvimento Cient\'{\i}fico e Tecnol\'ogico (CNPq), and Coordena\c{c}\~ao de Aperfei\c{c}oamento de Pessoal de N\'ivel Superior (CAPES).

\end{acknowledgments}

\bibliographystyle{h-physrev}
\bibliography{references}
\end{document}